\RequirePackage{lineno}
 \documentclass[aps,prd,preprint,showpacs,superscriptaddress,floatfix]{revtex4}

\usepackage{verbatim}
\usepackage{xspace}
\usepackage{color}
\usepackage{graphicx}% Include figure files
\usepackage{dcolumn}% Align table columns on decimal point
\usepackage{bm}% bold math
\usepackage{epstopdf}
\usepackage[%colorlinks% 
  bookmarks%
  ,pdftitle={DijetMassSpectra}%
  ,pdfsubject={CDF Run 2 Physics Result}%
  ,pdfauthor={CDF Collaboration}%
  ,pdfstartview={Fit}%
  ]{hyperref}
\usepackage[all]{hypcap}
\usepackage{multirow}
\usepackage{subfig}
\DeclareGraphicsExtensions{.pdf,.eps,.png,.jpg,.jpeg}
\graphicspath{{figures/}}

\usepackage{caption}
\usepackage{slashed}

\newcommand{\ppbar}{\ensuremath{p\overline{p}}\xspace}

\def\mMEt{\not\kern-.35em {E_T}}

\newcommand{\mett}{\mbox{$E\!\!\!\!/_{T}$}\xspace}

\newcommand{\met}{\mett}

\newcommand{\invfb}{\ensuremath{\rm fb^{-1}}\,}
\def\simle{\mathrel{
   \rlap{\raise 0.511ex \hbox{$<$}}{\lower 0.511ex \hbox{$\sim$}}}}

\begin{document}
%%%%%%%%%%%%%%%%%%%%%%%%%%%%%%%%%%%%%%%%%%%%%%%
% Toggle double spacing and line numbering
% Won't work with the PRD revtex4 !
%\pagewiselinenumbers % collab
%\linenumbers
%\doublespace
%%%%%%%%%%%%%%%%%%%%%%%%%%%%%%%%%%%%%%%%%%%%%%%

%\mbox{FERMILAB-PUB-XX-YYY-Z}
%\preprint{FERMILAB-PUB-11-375-E-PPD}
%\preprint{CDF/PHYS/ELECTROWEAK/PUBLIC/xxxx}

%\include{phys_defs.tex} 

\title{Invariant-mass distribution of jet pairs produced in
  association with a $W$ boson in \ppbar collisions at
  $\sqrt{s}=1.96$ TeV using the full CDF Run II data set}

%\author{The CDF Collaboration}
%\affiliation{\url{http://www-cdf.fnal.gov}}

%\date{February 17, 2014}

% Last update: $Date: 2014/02/17 18:16:27 $
\affiliation{Institute of Physics, Academia Sinica, Taipei, Taiwan 11529, Republic of China}
\affiliation{Argonne National Laboratory, Argonne, Illinois 60439, USA}
\affiliation{University of Athens, 157 71 Athens, Greece}
\affiliation{Institut de Fisica d'Altes Energies, ICREA, Universitat Autonoma de Barcelona, E-08193, Bellaterra (Barcelona), Spain}
\affiliation{Baylor University, Waco, Texas 76798, USA}
\affiliation{Istituto Nazionale di Fisica Nucleare Bologna, \ensuremath{^{ii}}University of Bologna, I-40127 Bologna, Italy}
\affiliation{University of California, Davis, Davis, California 95616, USA}
\affiliation{University of California, Los Angeles, Los Angeles, California 90024, USA}
\affiliation{Instituto de Fisica de Cantabria, CSIC-University of Cantabria, 39005 Santander, Spain}
\affiliation{Carnegie Mellon University, Pittsburgh, Pennsylvania 15213, USA}
\affiliation{Enrico Fermi Institute, University of Chicago, Chicago, Illinois 60637, USA}
\affiliation{Comenius University, 842 48 Bratislava, Slovakia; Institute of Experimental Physics, 040 01 Kosice, Slovakia}
\affiliation{Joint Institute for Nuclear Research, RU-141980 Dubna, Russia}
\affiliation{Duke University, Durham, North Carolina 27708, USA}
\affiliation{Fermi National Accelerator Laboratory, Batavia, Illinois 60510, USA}
\affiliation{University of Florida, Gainesville, Florida 32611, USA}
\affiliation{Laboratori Nazionali di Frascati, Istituto Nazionale di Fisica Nucleare, I-00044 Frascati, Italy}
\affiliation{University of Geneva, CH-1211 Geneva 4, Switzerland}
\affiliation{Glasgow University, Glasgow G12 8QQ, United Kingdom}
\affiliation{Harvard University, Cambridge, Massachusetts 02138, USA}
\affiliation{Division of High Energy Physics, Department of Physics, University of Helsinki, FIN-00014, Helsinki, Finland; Helsinki Institute of Physics, FIN-00014, Helsinki, Finland}
\affiliation{University of Illinois, Urbana, Illinois 61801, USA}
\affiliation{The Johns Hopkins University, Baltimore, Maryland 21218, USA}
\affiliation{Institut f\"{u}r Experimentelle Kernphysik, Karlsruhe Institute of Technology, D-76131 Karlsruhe, Germany}
\affiliation{Center for High Energy Physics: Kyungpook National University, Daegu 702-701, Korea; Seoul National University, Seoul 151-742, Korea; Sungkyunkwan University, Suwon 440-746, Korea; Korea Institute of Science and Technology Information, Daejeon 305-806, Korea; Chonnam National University, Gwangju 500-757, Korea; Chonbuk National University, Jeonju 561-756, Korea; Ewha Womans University, Seoul, 120-750, Korea}
\affiliation{Ernest Orlando Lawrence Berkeley National Laboratory, Berkeley, California 94720, USA}
\affiliation{University of Liverpool, Liverpool L69 7ZE, United Kingdom}
\affiliation{University College London, London WC1E 6BT, United Kingdom}
\affiliation{Centro de Investigaciones Energeticas Medioambientales y Tecnologicas, E-28040 Madrid, Spain}
\affiliation{Massachusetts Institute of Technology, Cambridge, Massachusetts 02139, USA}
\affiliation{University of Michigan, Ann Arbor, Michigan 48109, USA}
\affiliation{Michigan State University, East Lansing, Michigan 48824, USA}
\affiliation{Institution for Theoretical and Experimental Physics, ITEP, Moscow 117259, Russia}
\affiliation{University of New Mexico, Albuquerque, New Mexico 87131, USA}
\affiliation{The Ohio State University, Columbus, Ohio 43210, USA}
\affiliation{Okayama University, Okayama 700-8530, Japan}
\affiliation{Osaka City University, Osaka 558-8585, Japan}
\affiliation{University of Oxford, Oxford OX1 3RH, United Kingdom}
\affiliation{Istituto Nazionale di Fisica Nucleare, Sezione di Padova, \ensuremath{^{jj}}University of Padova, I-35131 Padova, Italy}
\affiliation{University of Pennsylvania, Philadelphia, Pennsylvania 19104, USA}
\affiliation{Istituto Nazionale di Fisica Nucleare Pisa, \ensuremath{^{kk}}University of Pisa, \ensuremath{^{ll}}University of Siena, \ensuremath{^{mm}}Scuola Normale Superiore, I-56127 Pisa, Italy, \ensuremath{^{nn}}INFN Pavia, I-27100 Pavia, Italy, \ensuremath{^{oo}}University of Pavia, I-27100 Pavia, Italy}
\affiliation{University of Pittsburgh, Pittsburgh, Pennsylvania 15260, USA}
\affiliation{Purdue University, West Lafayette, Indiana 47907, USA}
\affiliation{University of Rochester, Rochester, New York 14627, USA}
\affiliation{The Rockefeller University, New York, New York 10065, USA}
\affiliation{Istituto Nazionale di Fisica Nucleare, Sezione di Roma 1, \ensuremath{^{pp}}Sapienza Universit\`{a} di Roma, I-00185 Roma, Italy}
\affiliation{Mitchell Institute for Fundamental Physics and Astronomy, Texas A\&M University, College Station, Texas 77843, USA}
\affiliation{Istituto Nazionale di Fisica Nucleare Trieste, \ensuremath{^{qq}}Gruppo Collegato di Udine, \ensuremath{^{rr}}University of Udine, I-33100 Udine, Italy, \ensuremath{^{ss}}University of Trieste, I-34127 Trieste, Italy}
\affiliation{University of Tsukuba, Tsukuba, Ibaraki 305, Japan}
\affiliation{Tufts University, Medford, Massachusetts 02155, USA}
\affiliation{University of Virginia, Charlottesville, Virginia 22906, USA}
\affiliation{Waseda University, Tokyo 169, Japan}
\affiliation{Wayne State University, Detroit, Michigan 48201, USA}
\affiliation{University of Wisconsin, Madison, Wisconsin 53706, USA}
\affiliation{Yale University, New Haven, Connecticut 06520, USA}

\author{T.~Aaltonen}
\affiliation{Division of High Energy Physics, Department of Physics, University of Helsinki, FIN-00014, Helsinki, Finland; Helsinki Institute of Physics, FIN-00014, Helsinki, Finland}
\author{S.~Amerio\ensuremath{^{jj}}}
\affiliation{Istituto Nazionale di Fisica Nucleare, Sezione di Padova, \ensuremath{^{jj}}University of Padova, I-35131 Padova, Italy}
\author{D.~Amidei}
\affiliation{University of Michigan, Ann Arbor, Michigan 48109, USA}
\author{A.~Anastassov\ensuremath{^{v}}}
\affiliation{Fermi National Accelerator Laboratory, Batavia, Illinois 60510, USA}
\author{A.~Annovi}
\affiliation{Laboratori Nazionali di Frascati, Istituto Nazionale di Fisica Nucleare, I-00044 Frascati, Italy}
\author{J.~Antos}
\affiliation{Comenius University, 842 48 Bratislava, Slovakia; Institute of Experimental Physics, 040 01 Kosice, Slovakia}
\author{G.~Apollinari}
\affiliation{Fermi National Accelerator Laboratory, Batavia, Illinois 60510, USA}
\author{J.A.~Appel}
\affiliation{Fermi National Accelerator Laboratory, Batavia, Illinois 60510, USA}
\author{T.~Arisawa}
\affiliation{Waseda University, Tokyo 169, Japan}
\author{A.~Artikov}
\affiliation{Joint Institute for Nuclear Research, RU-141980 Dubna, Russia}
% ...

\collaboration{CDF Collaboration}
\altaffiliation[With visitors from]{
\ensuremath{^{a}}University of British Columbia, Vancouver, BC V6T 1Z1, Canada,
\ensuremath{^{b}}Istituto Nazionale di Fisica Nucleare, Sezione di Cagliari, 09042 Monserrato (Cagliari), Italy,
\ensuremath{^{c}}University of California Irvine, Irvine, CA 92697, USA,
\ensuremath{^{d}}Institute of Physics, Academy of Sciences of the Czech Republic, 182~21, Czech Republic,
\ensuremath{^{e}}CERN, CH-1211 Geneva, Switzerland,
\ensuremath{^{f}}Cornell University, Ithaca, NY 14853, USA,
\ensuremath{^{g}}University of Cyprus, Nicosia CY-1678, Cyprus,
\ensuremath{^{h}}Office of Science, U.S. Department of Energy, Washington, DC 20585, USA,
\ensuremath{^{i}}University College Dublin, Dublin 4, Ireland,
\ensuremath{^{j}}ETH, 8092 Z\"{u}rich, Switzerland,
\ensuremath{^{k}}University of Fukui, Fukui City, Fukui Prefecture, Japan 910-0017,
\ensuremath{^{l}}Universidad Iberoamericana, Lomas de Santa Fe, M\'{e}xico, C.P. 01219, Distrito Federal,
\ensuremath{^{m}}University of Iowa, Iowa City, IA 52242, USA,
\ensuremath{^{n}}Kinki University, Higashi-Osaka City, Japan 577-8502,
\ensuremath{^{o}}Kansas State University, Manhattan, KS 66506, USA,
\ensuremath{^{p}}Brookhaven National Laboratory, Upton, NY 11973, USA,
\ensuremath{^{q}}Queen Mary, University of London, London, E1 4NS, United Kingdom,
\ensuremath{^{r}}University of Melbourne, Victoria 3010, Australia,
\ensuremath{^{s}}Muons, Inc., Batavia, IL 60510, USA,
\ensuremath{^{t}}Nagasaki Institute of Applied Science, Nagasaki 851-0193, Japan,
\ensuremath{^{u}}National Research Nuclear University, Moscow 115409, Russia,
\ensuremath{^{v}}Northwestern University, Evanston, IL 60208, USA,
\ensuremath{^{w}}University of Notre Dame, Notre Dame, IN 46556, USA,
\ensuremath{^{x}}Universidad de Oviedo, E-33007 Oviedo, Spain,
\ensuremath{^{y}}CNRS-IN2P3, Paris, F-75205 France,
\ensuremath{^{z}}Universidad Tecnica Federico Santa Maria, 110v Valparaiso, Chile,
\ensuremath{^{aa}}The University of Jordan, Amman 11942, Jordan,
\ensuremath{^{bb}}Universite catholique de Louvain, 1348 Louvain-La-Neuve, Belgium,
\ensuremath{^{cc}}University of Z\"{u}rich, 8006 Z\"{u}rich, Switzerland,
\ensuremath{^{dd}}Massachusetts General Hospital, Boston, MA 02114 USA,
\ensuremath{^{ee}}Harvard Medical School, Boston, MA 02114 USA,
\ensuremath{^{ff}}Hampton University, Hampton, VA 23668, USA,
\ensuremath{^{gg}}Los Alamos National Laboratory, Los Alamos, NM 87544, USA,
\ensuremath{^{hh}}Universit\`{a} degli Studi di Napoli Federico I, I-80138 Napoli, Italy
}
\noaffiliation
% Last update: $Date: 2014/02/17 18:16:27 $
\begin{abstract}
  % remove the space for publication
   %This is the abstract
We report on a study of the dijet invariant-mass distribution in events with one identified lepton, a significant imbalance in the total event transverse momentum, and two jets. This distribution is sensitive to the possible production of a new particle in association with a $W$ boson, where the boson decays leptonically. We use the full data set of proton-antiproton collisions at 1.96 TeV center-of-mass energy collected by the Collider Detector at the Fermilab Tevatron and corresponding to an integrated luminosity of 8.9 fb$^{-1}$. The data are found to be consistent with standard-model expectations, and a 95$\%$ confidence level upper limit is set on the cross section for a $W$ boson produced in association with a new particle decaying into two jets. 
	
\end{abstract}

\maketitle

\section{Introduction}
\label{sec:introduction}
At hadron colliders the production of jets in association with vector bosons allows for precision tests of combined electroweak and quantum-chromodynamic (QCD) theoretical predictions. Many extensions of the standard model (SM) predict significant deviations from the SM predictions of the observable phenomena associated with these processes~\cite{hagiwara,kober,technicolor}.  In a previous publication, we reported a disagreement between data and SM expectations in a data sample corresponding to 4.3~\invfb\ \cite{bumpprl}. This disagreement appeared as an excess of events in the 120-160 GeV/$c^2$ invariant-mass range of the jet pairs ($M_{jj}$) for events selected by requiring one identified lepton, an imbalance in the total event transverse momentum, and two jets. 
Assuming that the excess of events over the SM prediction was due to an unknown contribution, modeled as a Gaussian resonance with width compatible with the expected dijet-mass resolution, the statistical significance of the excess was 3.2 standard deviations.  Similar searches carried out by the D\O\ \cite{Abazov:2011af}, CMS~\cite{WjjCMS}, and ATLAS~\cite{WjjATLAS} collaborations did not confirm the CDF result in events with the same topology. Another search for a dijet resonance carried out by the CDF collaboration in events with large missing transverse energy and two or three jets observed good agreement between data and SM expectations~\cite{METjjCDF}.

In this paper, we report on an update of the previous analysis~\cite{bumpprl} using the full CDF Run~II data set, which corresponds  to more than doubling the candidate event sample. In addition to the larger data set, we investigate in more detail a number of additional systematic effects. As a result of these studies, improved calibrations of detector response and modeling of instrumental backgrounds are used, yielding better agreement between data and SM expectations as obtained from Monte Carlo (MC) event generators. By incorporating these improved models, we perform a search for an excess of events over SM expectations in the dijet mass spectrum equivalent to the search described in Ref.\ \cite{bumpprl}.  %No significant excess is observed and a limits is placed on the cross section of a $W$ boson produced in association with a heavy resonance decaying into two jets.

The paper is structured as follows. In Sec.~\ref{sec:Detector} we describe the CDF II detector and the reconstruction of the final-state particles. In Sec. \ref{sec:JESModeling} we describe the independent energy corrections for simulated quark and gluon jets. In Sec.~\ref{sec:selection} we describe the candidate event selection and the expected composition of the sample. The background modeling is described in Sec.~\ref{sec:SignalBackgroundModeling}. The fitting method used in the analysis is described in Sec.~\ref{sec:Fit_Techniques}, and the results are given in Sec.~\ref{sec:Results}. We discuss the conclusions in Sec.~\ref{sec:conclusion}.

More information about the studies reported in this paper can be found in Ref.\ \cite{MarcoTrovatoThesis}.

% The implications of these systematic uncertainty studies on physics measurements in jets at hadron colliders is discussed.

%%% Local Variables: 
%%% mode: latex
%%% TeX-master: "DijetMassSpectra.tex"
%%% End: 

\section{Event Detection and Reconstruction}
\label{sec:Detector}
Details on the CDF~II detector and the event reconstruction are described elsewhere \cite{CDF_detect_A}.  The detector is cylindrically symmetric around the $z$ direction, which is oriented along the proton beam axis.  The polar angle, $\theta$, is measured from the origin of the coordinate system at the center of the detector with respect to the $z$ axis. Pseudorapidity, transverse energy, and transverse momentum are defined as $\eta$=$-\ln\tan(\theta/2)$, $E_{T}$=$E\sin\theta$, and $p_{T}$=$p\sin\theta$ respectively, where $E$ is the energy measured in a calorimeter tower (or related to an energy cluster) with centroid at angle $\theta$ with respect to the nominal collision point, and $p$ is a charged-particle momentum. The azimuthal angle is labeled $\phi$. Trajectories of charged particles (tracks) are determined using a tracking system immersed in a 1.4~T magnetic field, aligned coaxially with the $\ppbar$ beams.  A silicon microstrip detector provides tracking over the radial range 1.5 to 28~cm.  A 3.1~m long open-cell drift chamber, the Central Outer Tracker (COT), covers the radial range from 40 to 137~cm and provides up to 96 measurements. Sense wires are arranged in eight alternating axial and $\pm2^{\circ}$ stereo ``superlayers'' with 12 wires each.  The fiducial
region of the silicon detector extends to $|\eta| \approx 2$, while the COT provides full coverage for $|\eta|\simle 1$. The momentum resolution for charged particles in the COT is $\delta p_T /p^2_T \approx 0.0015$, where $p_T$ is in units of GeV/$c$. The central and plug calorimeters, which cover the pseudorapidity regions of $|\eta|<$ 1.1 and 1.1 $<|\eta|<$ 3.6 respectively, are divided into a front electromagnetic and a rear hadronic compartment, which surround the tracking system in a projective-tower geometry. Muons with $|\eta| <$ 1 are detected by drift chambers and scintillation counters located outside the hadronic calorimeters.

%The transverse energy is measured in each calorimeter tower where the polar angle $\theta$ is calculated using the measured z position of the event vertex and the tower location.
Contiguous groups of calorimeter towers with signals exceeding a preset minimum are identified and summed together into energy clusters. An electron candidate, referred to as a ``tight central electron'', is identified in the central electromagnetic calorimeter as an isolated, mostly electromagnetic cluster matched to a reconstructed track in the pseudorapidity range $|\mathrm\eta|<$ 1.1. The electron transverse energy is reconstructed from the
electromagnetic cluster with an uncertainty $\sigma(E_T )/E_T \approx
13.5\%/\sqrt{E_T~(\rm GeV)} \oplus 1.5\%$. 

A hadron jet is identified as a cluster of calorimeter energies contained within a cone of radius $\Delta R \equiv \sqrt{(\Delta \phi)^2 + (\Delta \eta)^2} = 0.4$, where $\Delta \eta$ and $\Delta \phi$ are the distances in pseudorapidity and azimuthal angle between a tower center and the cluster axis. Jet energies are corrected for a number of effects that bias the measurement \cite{cdf_JES}. These corrections include imposing uniformity of calorimeter response as a function of $|\eta|$, removing expected contributions from multiple \ppbar interactions per bunch crossing, and accounting for nonlinear response of the calorimeters. % The jet energy resolution is approximately $\sigma(E_T ) \approx [0.1E_T + 1.0 GeV]$. 
These corrections are applied generically to all reconstructed jets independent of the flavor of the associated parton, which is responsible for initiating the particle shower. Recent studies demonstrate the need for additional corrections to the reconstructed energies of jets in simulated events dependent on the flavor of the initiating parton in order to correctly model the observed energy scale of reconstructed jets in data\ \cite{Diboson_lljj}.  These additional corrections, applied in the analysis described here, are discussed in greater detail in Sec.\ \ref{sec:JESModeling}.
%The standard CDF jet-energy corrections do not account for potential differences in calorimeter response to jets originated from gluons or quarks. In Sec.\ \ref{sec:JESModeling} we describe additional energy corrections to simulated quark and gluon jets that are applied in this analysis \cite{Diboson_lljj}.

Muons are identified in three independent subdetectors. Muons with $|\eta| \leq$ 0.6 and $p_T > 1.4~{\rm GeV}/c$ are detected in four layers of planar drift chambers (CMU) located outside the central calorimeter at five interaction lengths. Muons with $|\eta| \leq$ 0.6  and $p_T>2.8~{\rm GeV}/c$ are detected in four additional layers of drift chambers (CMP) located at eight interaction lengths of calorimeter and steel absorber.  Muons with $0.6 \leq |\eta| \leq 1.0$  and $p_T > 2.2~{\rm GeV}/c$ are detected by a system of eight layers of drift chambers and scintillation counters (CMX) located outsied the calorimeter at six to ten absorption lengths.  Muon candidates are identified by extrapolating isolated tracks to track segments in the muon detector systems.

Missing transverse energy (\met) is defined as the magnitude of the vector sum of all calorimeter-tower energy depositions projected on the transverse plane. It is used as a measure of the sum of the transverse momenta of the particles that escape detection, most notably neutrinos. The vector sum includes corrected jet energies and also the momenta of high-$p_T$ muon candidates, which deposit only a small fraction of their energy in the calorimeter.

\section{Quark and gluon energy scale modeling}
\label{sec:JESModeling}
The modeling of calorimeter response to particle showers originating from quarks and gluons is dependent on the different fragmentation and hadronization models used in the simulation for each.  Hence, the level of agreement between the simulated and observed energy scales of jets originating from quarks and gluons can differ significantly. We derive specific corrections for the calorimeter response to quark and gluon jets in simulated events using two independent samples of jets with different quark fraction. We use one sample where a jet is emitted in an opposite direction with respect to an energetic photon in the transervse plane, and another sample of $Z \rightarrow \ell^{+}\ell^{-} + $ jet events ($\ell$ being an electron or muon). The former sample is richer in quark jets, the latter in gluon jets. Photon and $Z$-boson energies are measured more accurately than jet energies and can be used to calibrate the jet energy as described below. The criteria for selecting events with a photon or $Z$ boson associated with only one jet are described in Ref.\ \cite{Diboson_lljj}. 

We derive independent corrections for the quark and gluon jet-energy scales in data and simulation through $Z+$jet and $\gamma+$jet samples. We define the jet-balance in $Z$+jet or $\gamma$+jet events as follows:

\begin{equation}
K_{Z,\gamma} = ( E_{T}\textsuperscript{jet} / p_{T}^{Z,\gamma}) - 1\text{.}
\label{eq:balance_def}
\end{equation}
The measured average balance is corrected with a jet-energy correction factor of $1/(K_{Z,\gamma} + 1)$.

The jet balance in Eq.\ (\ref{eq:balance_def}) can be rewritten as the weighted average of the balance variables for quark and gluon jets, $K_{q}$ and $K_{g}$ respectively. If $F_{X}^{q,g}$ is the quark, or gluon fraction in sample $X$, then we write

\begin{equation}
K_{Z} = F_{Z}^{q}K_{q} + F_{Z}^{g}K_{g} = F_{Z}^{q}K_{q} + (1 - F_{Z}^{q})K_{g}
\label{eq:balance_Z}
\end{equation}
\begin{equation}
K_{\gamma} = F_{\gamma}^{q}K_{q} + F_{\gamma}^{g}K_{g} = F_{\gamma}^{q}K_{q} + (1 - F_{\gamma}^{q})K_{g}\text{,} \label{eq:balance_gamma}
\end{equation}
or, solving for $K_{q}$ and $K_{g}$,

\begin{equation}
K_{q} = \frac{1}{F_{\gamma}^{q} - F_{Z}^{q}} [ (1 - F_{Z}^{q})K_{\gamma} - (1 - F_{\gamma}^{q})K_{Z} ]
\label{eq:balance_q}
\end{equation}
\begin{equation}
K_{g} = \frac{1}{F_{\gamma}^{q} - F_{Z}^{q}} [ F_{\gamma}^{q}K_{Z} - F_{Z}^{q}K_{\gamma} ]\text{.}
\label{eq:balance_g}
\end{equation}
These equations apply separately to data and MC simulation with distinct balance factors $K_{X}\textsuperscript{d}$ and $K_{X}\textsuperscript{MC}$ and can include a dependence on the energy of the jet, $F_{X}^{q} \rightarrow F_{X}^{q}(E_{T}\textsuperscript{jet})$ and $K_{X} \rightarrow K_{X}(E_{T}\textsuperscript{jet})$.

In order to solve for $K_{q}$ and $K_{g}$, we need to input the values of $K_{Z,\gamma}$ and $F_{Z,\gamma}^{q}$. We extract the former in data and simulation by constructing the balancing distribution, as defined in Eq.\ (\ref{eq:balance_def}), in bins of $E_{T}\textsuperscript{jet}$, and fitting the core of the distribution around its maximum with a Gaussian function. %The mean and uncertainty on the mean of the fitted Gaussian are used respectively as the value of $K_{Z,\gamma}(E_{T}\textsuperscript{jet})$ and as its uncertainty. The Gaussian fit is performed in both data and simulations. 
We determine $F_{Z,\gamma}^{q}$ in simulation by matching jets to their originating partons, by requiring $\Delta R<0.4$ between the parton and the jet. In the $\gamma +$jet balancing sample the quark fraction is about $85\%$ at $E_{T}\textsuperscript{jet} \approx 30$~GeV, and reduces to about $71\%$ at $E_{T}\textsuperscript{jet} \approx 70$~GeV. In the $Z +$jet balancing sample this fraction is about $38\%$ and $49\%$ in the same $E_{T}\textsuperscript{jet}$ ranges.
In data, it is not possible to match jets to their originating parton, and we rely on the values of $F_{Z,\gamma}^{q}(E_{T}\textsuperscript{jet})$ extracted from the simulated samples. %Because we are trying to correct for discrepancies in the reconstruction of quark and gluon jets between data and simulation, we cannot simply use the simulation-derived $F_{Z,\gamma}^{q}$ values from each jet $E_{T}$ bin. Instead, we can use the reconstructed $p_{T}^{Z,\gamma}$, which should match in data and simulation, and determine the quark/gluon fraction based on that value. Therefore, we parametrize the $F_{Z,\gamma}^{q}$ from simulation as a function of $p_{T}^{Z,\gamma}$%:
%$$F_{Z/\gamma}^{q}\textsuperscript{MC}(p_{T}) = a + e^{bp_{t} + c}$$
%and determine the $F_{Z/\gamma}^{q}\textsuperscript{data}$ in each jet $E_{T}$ of the data based on $p_{T}^{Z/\gamma}$ distribution in the data.

\begin{figure*}[htbp]
\centering
{\includegraphics[width=0.8\textwidth]{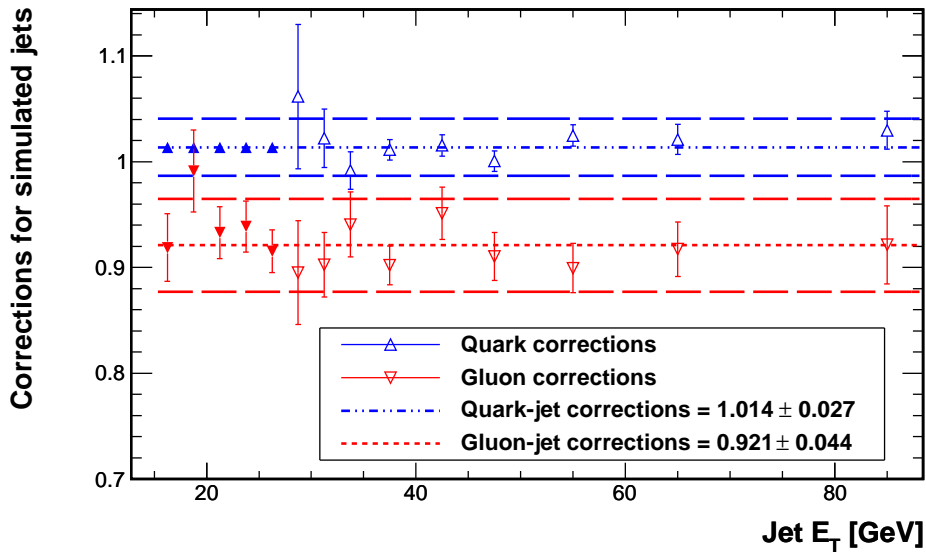}}
\captionsetup{labelsep=period}
\caption{Derived energy scale corrections for simulated quark jets and gluon jets as a function of jet $E_T$. The open triangles represent corrections derived jointly from the $\gamma +$jet and $Z +$jet balancing samples, while the filled triangles in the low-jet $E_T$ region are obtained from the $Z +$jet sample only, assuming a constant correction for the quark jet-energy scale. Error bars are from statistical sources only. The short dashed lines show the fits to constant energy corrections, and the long dashed lines represent the total systematic uncertainty bands on the correction determined by the fit.}
\label{fig:JES_correction}
\end{figure*}

Using Eqs.\ (\ref{eq:balance_q})-(\ref{eq:balance_g}), we derive $K_{q}$ and $K_{g}$ in data and simulation as functions of jet $E_T$. Rather than correcting both data and simulation, the factors $K_{q}$ and $K_{g}$ are used to determine the corrections to simulated jets, in order to best match the energy scale observed in data. These corrections are defined as $(K_{q}\textsuperscript{d} + 1)/(K_{q}\textsuperscript{MC} + 1)$ for quark jets and $(K_{g}\textsuperscript{d} + 1)/(K_{g}\textsuperscript{MC} + 1)$ for gluon jets, the extracted values for which are shown in Fig.\ \ref{fig:JES_correction}. 

The transverse energy threshold of the photon online event-selection (trigger) is 25 GeV \cite{Abe:1993qb}, so reliable balancing information is not available for jets with energies less than 27.5 GeV in the photon-triggered sample. Since we are interested in jets with energies extending down to 20 GeV, we extrapolate the quark-jet-energy corrections to lower jet energies, and use the $Z +$jet balancing sample to extract a gluon correction assuming this extrapolated quark correction.
\begin{comment}
\begin{figure*}[htbp]
\centering
{\includegraphics[width=0.49\textwidth]{balance_quark.eps}}
{\includegraphics[width=0.49\textwidth]{balance_gluon.eps}}
\caption{The derived balancing variable for quark jets, $K_{q}$, (\textit{left}) and gluon jets, $K_{g}$, (\textit{right}) in data (\textit{black}) and simulation (\textit{red}) as a function of  $E_T\textsuperscript{jet}$. The uncertainties on each point are from the uncertainties from the mean of the Gaussian fit and the uncertainties on the quark fractions, added in quadrature. We see better agreement between data and simulation in the energy scale of quark jets than that of gluon jets, following from the behavior seen in Fig.\ \ref{fig:balance_z-gamma}.}
\label{fig:balance_q-g}
\end{figure*}
\end{comment}

As both the quark and gluon corrections do not depend on jet energy for jets with $E_{T} \ge 15$~GeV, we fit them to a constant. To better match the data, quark-jet energies in the simulation should be increased by $(1.4 \pm 2.7)\%$, while gluon-jet energies should be decreased by $(7.9 \pm 4.4) \%$. The reported uncertainties are the sum in quadrature of the statistical and systematic contributions. The systematic sources are dominated by a 10$\%$ uncertainty on the quark fractions in the $Z+$jet or $\gamma +$jet balancing samples. The uncertainty is estimated by fitting the data distribution of a quark-gluon discriminant parameter \cite{Diboson_lljj} with quark and gluon templates from simulation. The average deviation of the extracted quark fraction from the prediction is taken as the systematic uncertainty on the quark fraction. Other sources of systematic uncertainties include the extrapolation to low quark-jet energy and the differences between the allowed number of interaction vertices in the $Z +$jet and $\gamma +$jet samples. The sizes of statistical and systematic uncertainties are comparable. Because of the default corrections applied to reconstructed jet energies, which are designed to equate the energy scales for simulated and observed jets on average, uncertainties on the additional, independent corrections derived for quark and gluon jets are necessarily anticorrelated with one another. Combination of these two anticorrelated uncertainties encompasses the uncertainty on the absolute energy scale for generic jets, which is the dominant uncertainty assigned to the default CDF jet-energy corrections.  In order to avoid double-counting, only the anticorrelated uncertainties associated with the additional quark and gluon corrections are applied within this analysis. The observation that the additional energy-scale correction
for quark jets is consisten with unity within measurement uncertainties is consistent with the {\it in situ} calibration of light-quark jet energies performed in conjunction with the top-quark mass measurement \cite{TopMassinsitu}.
%Because the corrections shift the relative energy response in simulation to match data, the quark-jet and gluon-jet-energy-correction uncertainties are anticorrelated. These uncertainties are used instead of those of the default CDF jet-energy-scale uncertainties~\cite{cdf_JES}. Both uncertainties are similar in magnitude. Moreover, the obtained correction of the quark-jet energies is compatible to the \textit{in situ} calibrations obtained in the top mass measurement \cite{TopMassinsitu}.

Similar studies in the $Z+$jet balancing sample show that the calorimeter responses to heavy-flavor quark jets in simulation and data agree. Since the uncertainty on the energy scale of heavy-quark jets relative to that of light-quark jets is roughly 1$\%$ \cite{TopMass}, possible discrepancies of the calorimeter responses to heavy-flavor quark jets in simulation and data are expected to be covered by the light-quark jet-energy-scale uncertainty.

%%% Local Variables: 
%%% mode: latex
%%% TeX-master: "DijetMassSpectra.tex"
%%% End: 

\section{Data Set and Event Selection}
\label{sec:selection}
We select a sample enriched in $W+$jets events by requiring a large transverse-momentum electron or muon passing the high-$p_T$ lepton trigger requirements, large missing transverse energy, and two energetic jets. The full CDF Run II data set is used, corresponding to an integrated luminosity of 8.9 fb$^{-1}$.

\subsection{Online event selection}
The trigger is a three-level event filter with tracking information available at the first level. The first level of the central-electron trigger requires a charged particle with $p_T>8~{\rm GeV}/c$ pointing to a calorimeter tower with $E^{\it EM}_T> 8~{\rm  GeV}$ and $E^{\it HAD} /E^{\it EM} < 0.125$, where $E^{HAD}$, $E^{EM}$ are the energy deposited by the candidate electron in the hadronic and electromagnetic calorimeters respectively. The first level of the muon trigger requires a charged particle with $p_T>4~{\rm GeV}/c$ or $8~{\rm GeV}/c$ pointing to a muon stub. Full lepton reconstruction (Sec.\ \ref{sec:Detector}) is performed at the third trigger level, with requirements of $E_T > 18~{\rm GeV}$ for central electrons and $p_T> 18~{\rm GeV}/c$ for muons.

\subsection{Offline event selection}

Offline, we select events containing exactly one electron with $E_T>20~{\rm GeV}$ or muon with $p_T>20~{\rm GeV}/c$, large missing transverse energy ($\met>25~{\rm GeV}$), and exactly two jets with $E_T>30~{\rm GeV}$ and $|\eta|<2.4$. In order to select events with $W$ bosons and to reject multijet backgrounds, we impose the following requirements: transverse mass $m_T>30~{\rm GeV}$, where $m_T=\sqrt{2  p_T^{\ell} \met \{1-\cos [\Delta \phi( {\vec{p}_T}^{\phantom{1}l},\vec{\met})]\}}$, $\ell$ being an electron or a muon; azimuthal angle between \met and the most energetic jet $\Delta \phi(\met,j_1)>0.4$; difference in pseudorapidity between the two jets $|\Delta \eta(j_1,j_2)|<2.5$; and transverse momentum of the dijet system $p_T^{jj}>40~{\rm GeV}/c$.  The position of the primary interaction is found by fitting a subset of well-measured tracks pointing to the beam line and is required to lie within 60\,cm from the center of the detector. If multiple vertices are reconstructed, the vertex associated with charged particles yielding the maximum scalar sum $p_T$ is defined as the primary-interaction point. The longitudinal coordinate $z_0$ of the lepton track at the point of closest approach to the beam line must also lie within 5\,cm of the primary-interaction point.

%%% Local Variables: 
%%% mode: latex
%%% TeX-master: "DijetMassSpectra.tex"
%%% End: 

%\includeorinput{JES_modeling}

\section{Signal and Background Modeling} 
\label{sec:SignalBackgroundModeling} 
We search for an excess of events in the invariant-mass spectrum of the two reconstructed jets from the decay of a potential non-SM particle. To be consistent with Ref.\ \cite{bumpprl}, we model the excess with a Gaussian function centered at a mass of 144 GeV/$c^2$ with a width of 14.3 GeV/$c^2$, determined by the calorimeter resolution expected from simulation.

There are two main categories of background processes: physics processes, such as the dominant $W+$jets mechanism, where all final-state particles are correctly identified, and instrumental background, where the lepton is misidentified and the missing transverse energy is mismeasured. The expected rates of the major backgrounds for a 20-300 GeV/$c^2$ dijet-mass range are reported in Table \ref{tab:RatesBumpCuts}, as obtained from the modeling of each background described below.

\begin{table*}[!ht]
\centering                          % used for centering table 
\captionsetup{labelsep=period}
\caption[]{Expected number of events in the 20-300 GeV/$c^2$ dijet-mass range with electron and muon candidates in the selected sample from each of the background processes. The total expected number of events is constrained to be equal to the number of observed events, as described in Sec.\ \ref{sec:mult-backgr-model}. The reported uncertainties and the central values for the $W/Z+$jets contributions are obtained from the \met\ fit (Sec.\ \ref{sec:mult-backgr-model}). The uncertainties on the top-quark-pair contribution are derived from the experimental measurement\ \cite{CDFttbarcrosssection}, those on the single-top-quark and diboson contributions come from the theoretical cross sections \cite{singletopschannelxsec, singletoptchannelxsec, campbell}. The central value and the uncertainty for the QCD multijet process is obtained from the \met\ fit (Sec.\ \ref{sec:mult-backgr-model}).}%Vjets uncert from figureEventSelection/Results/resultfit_BuildTemplatesForSignificance_withSyst_bumpcutsetaless24_allimprovements_TCECMUPCMX_56bins20to300_Jan3
\begin{tabular}{ccc }
\hline
\hline
Production process\hspace{1cm} & Events (electron channel)\hspace{1cm} & Events (muon channel)\\
\hline 
$W+$jets & 8900 $\pm$ 119 & 5959 $\pm$ 95\\
$Z+$jets & 248 $\pm$ 3 & 472 $\pm$ 9\\
$t\overline{t}$ & 670 $\pm$ 44 & 431 $\pm$ 28\\
Single-top & 161 $\pm$ 10 & 106 $\pm$ 7 \\
Diboson & 589 $\pm$ 36  & 392 $\pm$ 24 \\
QCD multijets & 898 $\pm$ 127  & 20 $\pm$ 3\\
\hline
Total expected	& 11466 $\pm$ 185 & 7380 $\pm$ 109\\
\hline
\hline
\end{tabular}
\label{tab:RatesBumpCuts}
\end{table*}

\subsection{Physics backgrounds: $W/Z$+jets, top-quark, and diboson production}
The dominant contributing process to the selected sample is the associated production of $W$ bosons and jets. Another process with a non-zero contribution to the selected sample is $Z+$jets, where a lepton from the $Z$-boson decay is not detected. The predicted ratio between number of events with heavy-flavor and light-flavor jets in $W/Z+$jets processes is about 10$\%$. To study the effects of $W$+jets and $Z$+jets processes, events are generated using {\sc alpgen} \cite{Mangano:2002ea} interfaced with {\sc pythia} \cite{pythia} for parton showering and hadronization. Because of large uncertainties associated with the NLO calculations \cite{VjetsNorm}, the magnitude of $W$+jets and $Z$+jets contributions is obtained from a fit to the data, where the ratio of the $W+$jets cross section to $Z+$jets cross sections is constrained to 3.5 as predicted by theory. %From Frisch's paper (although the predictions are made by using MCFM. Uncertainty is negligible 0.06 as show from table iV CDF note 5963)
Top-quark pair production is modeled with events simulated using {\sc pythia} and assuming a top-quark mass of 172.5 GeV/$c^2$. The magnitude of the simulated top-pair contribution is normalized based on the latest CDF measurement on an independent sample with one identified lepton, significant transverse momentum imbalance, and at least three jets \cite{CDFttbarcrosssection}. The uncertainty of the top-quark pair cross section is 7\%.  Processes producing a single top quark are modeled by the {\sc madevent} event generator \cite{MADEVENT} interfaced to {\sc pythia} for showering and hadronization.  The cross sections are normalized to the next-to-next-leading order (NNLO) plus next-to-next leading log (NNLL) for the $s$-channel \cite{singletopschannelxsec} and next-to-next-to-next-leading order (NNNLO) plus next-to-leading log (NLL) for the $t$-channel \cite{singletoptchannelxsec} theoretical calculations, with uncertainties of 11\%.

Diboson ({\it WW}, {\it WZ}, {\it ZZ}) production is modeled with {\sc pythia}. Expected diboson contributions are normalized based on the theoretical NLO cross sections \cite{campbell}. The resulting uncertainty on the diboson contribution is roughly 6\%.

The remaining background process is multijet production, where one jet mimics the experimental signature of a lepton and a mismeasurement in the calorimeter leads to spurious \met in the event. We use data to model this contribution, as described in Sec.~\ref{sec:mult-backgr-model}.

\begin{figure*}[htbp]
\centering
{\includegraphics[width=.79\textwidth]{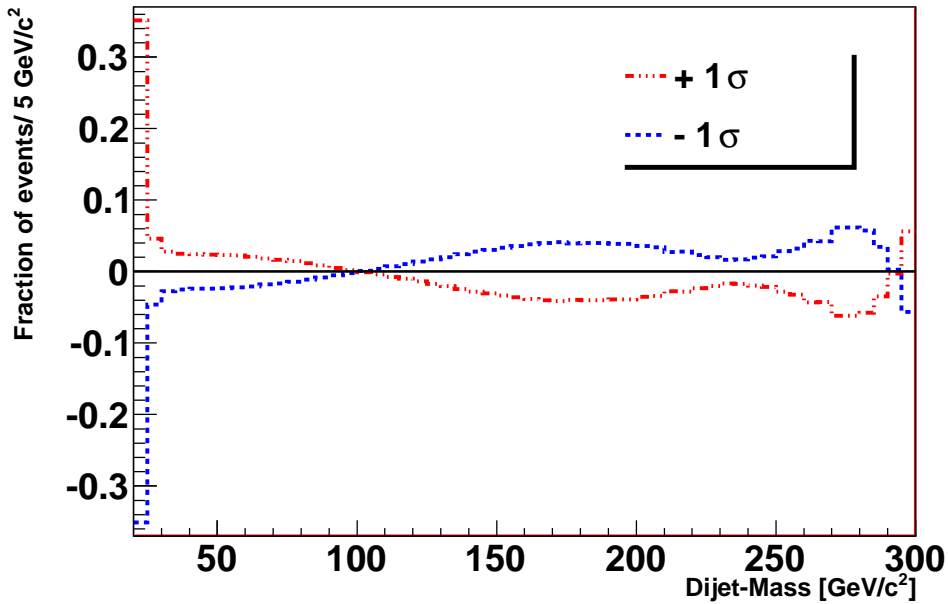}}
\captionsetup{labelsep=period}
\caption{Relative difference in the combined electron and muon samples between the nominal dijet mass distribution and the one obtained by varying the jet-energy scale by $\pm 1\sigma$ in $W$+jets events.}
\label{fig:jeshiVsjesloVsnominalWjetsBumpCuts}
\end{figure*}

Other sources of systematic uncertainties that affect the background normalizations are those associated with the luminosity measurement (6\%) \footnote{Since the magnitude of top-quark-pair contribution is effectively insensitive to the uncertainty on luminosity \cite{CDFttbarcrosssection}, the luminosity uncertainty has not been applied to this contribution.}, effects of initial-state and final-state radiation (2.5\%), modeling of the parton distribution functions (2.2\%), modeling of the jet-energy scale (2.7\% for quark jets and 4.4\% for gluon jets with a 100\% anticorrelation), modeling of the jet-energy resolution (0.7\%), and modeling of the trigger efficiency (2.2\%). In addition to uncertainties on the expected contributions from each background process, we also consider systematic uncertainties that affect the shape of the invariant-mass distribution for each process. The most important are the uncertainties on the jet-energy scale and on the renormalization and factorization scales in the $W$ + jets process, which are taken to be equal. For modeling the former, two alternative invariant-mass distributions are obtained by varying the jet-energy scale within its expected $\pm 1\sigma$ uncertainty. For the latter, the factorization scale used in the event generation \footnote{$Q^2 = M_W^2 + p_T^2$, where $M_W =$ 80.4 GeV/$c^2$ and $p_T^2$ is the squared sum of transverse energies of all final-state partons.} is doubled and halved in order to obtain two alternative shapes. As an example, the relative difference between the varied and nominal shapes for the dominant background process ($W+$jets) due to the jet-energy-scale variation is shown in Fig.\ \ref{fig:jeshiVsjesloVsnominalWjetsBumpCuts}.%-\ref{fig:q2downVsq2upVsnominalWjetsBumpCuts}.

\begin{comment}
\begin{figure}[htbp]
\centering
%\includegraphics[width=.49\textwidth]{figures/SystUncertainty/EleMuo/wjets_q2_downmnominal_pretag_TCECMUPCMX.eps}
\includegraphics[width=.49\textwidth]{figures/SystUncertainty/EleMuo/wjets_q2_downnominal_pretag_TCECMUPCMX_ratio.eps}
\captionsetup{labelsep=period}
\caption{Difference in the combined electron and muon samples between the nominal dijet mass distribution and the one obtained by halving the nominal $Q^2$ in $W$+jets events. The electron and muon samples have been combined. The difference between the nominal and the one obtained by doubling the nominal $Q^2$ is defined by symmetrizing each bin content with respect to zero.}
\label{fig:q2downVsq2upVsnominalWjetsBumpCuts}
\end{figure}
\end{comment}

\subsection{Multijet production} 
\label{sec:mult-backgr-model}
Multijet events can be identified as signal candidates when one of the jets is misidentified as a lepton. This mismeasurement can also result in significant missing transverse energy. Because it is unlikely for a jet to deposit energy in the muon detectors, the misidentification probability of a muon is lower than that of an electron. The multijet-background contribution is thus negligible in the muon channel ($<0.5$\%), while it is close to 10\% in the electron channel (Table \ref{tab:RatesBumpCuts}). Therefore, we concentrate on discussing the multijet-background modeling for events with electron candidates. Similar methods are used to model this background for muon events. %

\begin{figure*}[htbp]
\centering
{\includegraphics[width=0.79\textwidth]{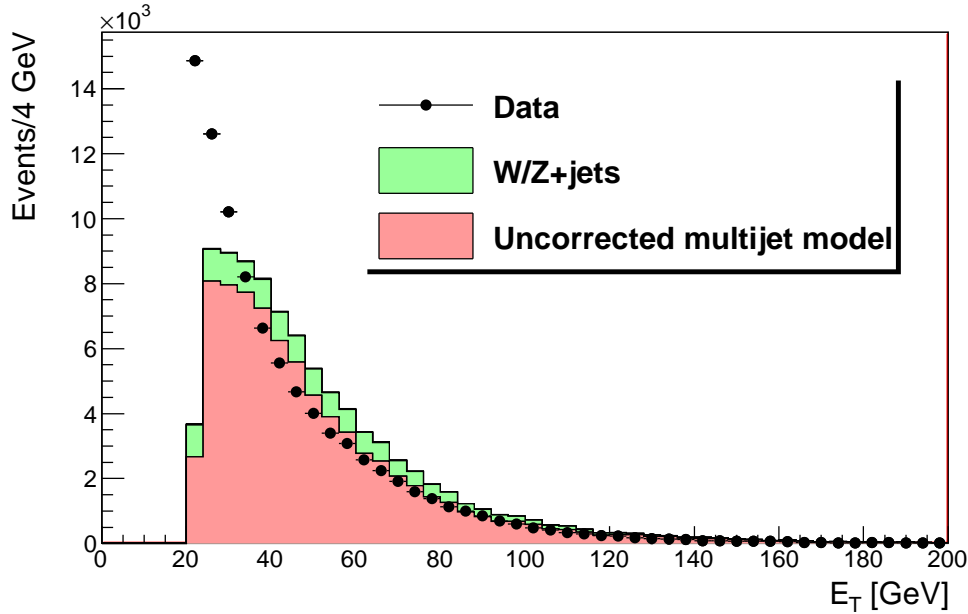}}
\captionsetup{labelsep=period}
\caption{Transverse-energy distribution of jets matched to identified electrons in the multijet-enriched control region in data (circles), uncorrected multijets model (dark shaded histogram), and $W/Z+$jets simulation (light shaded histogram). The magnitude of $W/Z$+jets contributions is normalized to the NLO calculations \cite{VjetsNorm}, while the magnitude of the multijet model is obtained from the data. In subsequent analysis, the multijet model is reweighted such that the predicted and observed energy spectra agree.}
\label{fig:jetmatched}
\end{figure*}

To model the multijet-background distribution, we use an event sample obtained from the same selection as described in Sec.~\ref{sec:Detector} except that two identification criteria for the electron candidates that do not depend on the kinematic of the event ({\it e.g.}, the fraction of energy in the hadronic calorimeter) are inverted \cite{Aaltonen:2010jr}. The particles identified with those inverted requirements are referred to as ``nonelectrons''. This ensures that the sample used for modeling the multijet background is statistically independent of the signal sample, while as similar as possible kinematically.
Nevertheless, several tunings are needed to this sample in order to adequately model the multijet component in the signal sample. First, there is a small contribution of events with prompt leptons from boson decays.  We subtract this contribution bin-by-bin for any variable of interest using the theoretical prediction for that bin. A second tuning of the nonelectron sample accounts for the trigger bias. The trigger selects events based on the $E_T$ of the reconstructed electron or nonelectron candidate, but the event kinematic properties are determined by the $E_T$ of the corresponding jet.  We define this jet as the jet with $\Delta R <$ 0.4 with respect to the (non)electron. To properly model the event kinematics properties, the energy distribution of this jet should be the same in events with misidentified electron and nonelectron candidates.  We define a control region enriched in multijet events, selected with the same criteria as for the signal region, except for the requirement of $\met <$ 20 GeV or $m_T <$ 30 GeV. The estimated fraction of multijet events in this region is $84\%$. When comparing the energy distribution of jets matched to misidentified electrons with jets matched to nonelectrons in this control region, we find discrepancies due to the trigger on electron $E_T$ (Fig.\ \ref{fig:jetmatched}). The jets matched to misidentified electrons have a higher fraction of their measured energy in the electromagnetic calorimeter than jets matched to nonelectrons; therefore, in order to have a nonelectron of the same energy as a corresponding misidentified electron, the energy of the jet producing the nonelectron must be higher. The trigger threshold thus leads to a higher average $E_T$ of jets producing nonelectrons than of jets producing misidentified electrons. To remove this trigger bias, we reweight events in the nonelectron sample such that the energy spectrum of the jets matched to misidentified electrons is equivalent to the energy spectrum of jets matched to nonelectrons. The reweighting is obtained from the control region and the same weights are used in the signal region.

A final tuning of the nonelectron sample addresses the difference in jet-energy scale between the jet producing the nonelectron and the jet producing a misidentified electron. We investigate this difference using {\sc pythia} QCD dijet events. For the same primary parton energy, the energy of jets matched to nonelectrons is systematically lower than the energy of jets matched to identified electrons. Based on the observed differences, we derive an energy correction factor as a function of the initial jet-energy, which is applied to events in the nonelectron sample.

\begin{figure*}[htbp]
\centering
\captionsetup{justification=centering}
\subfloat[] {\includegraphics[width=.49\textwidth]{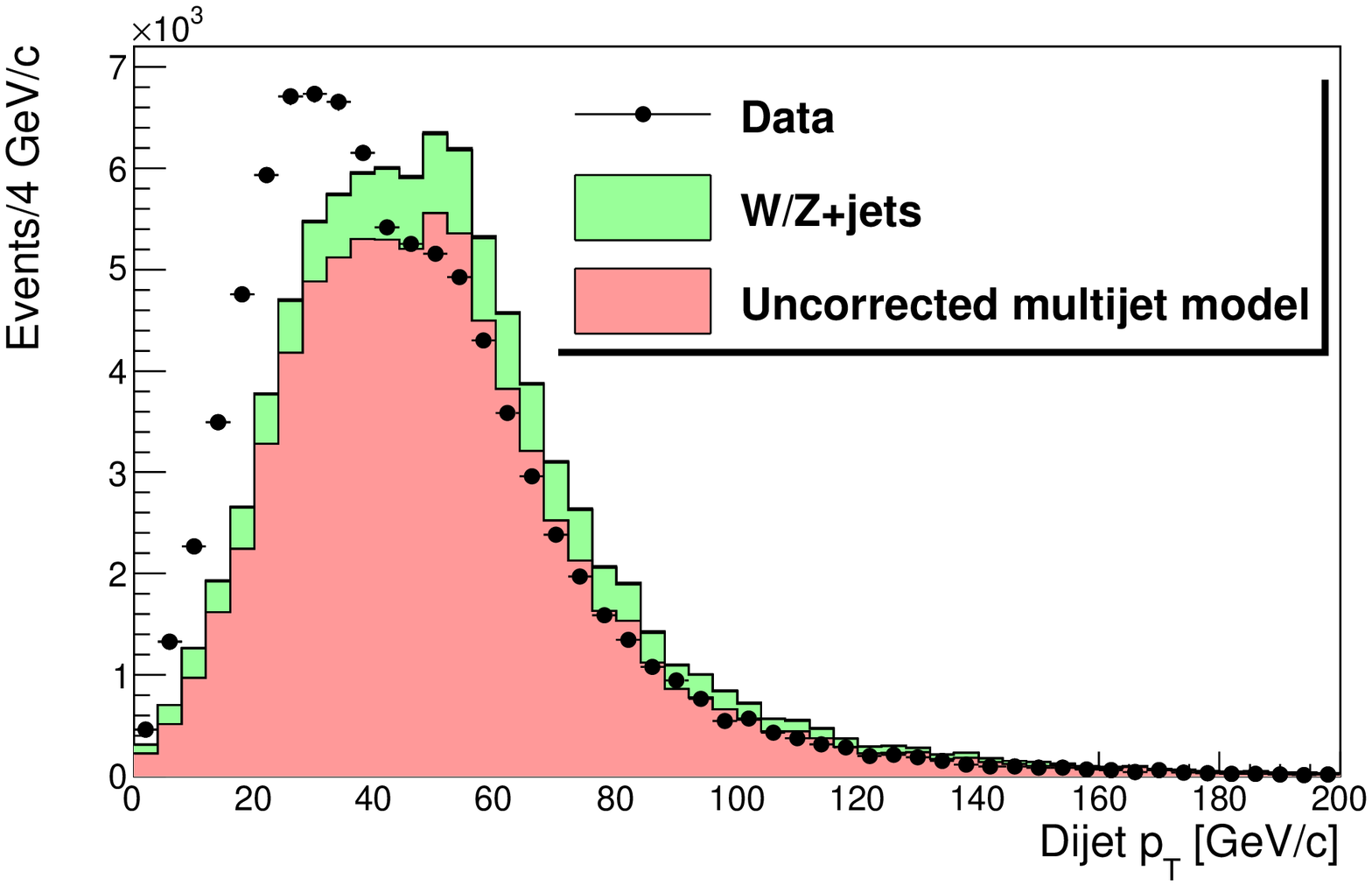}}
\subfloat[] {\includegraphics[width=.49\textwidth]{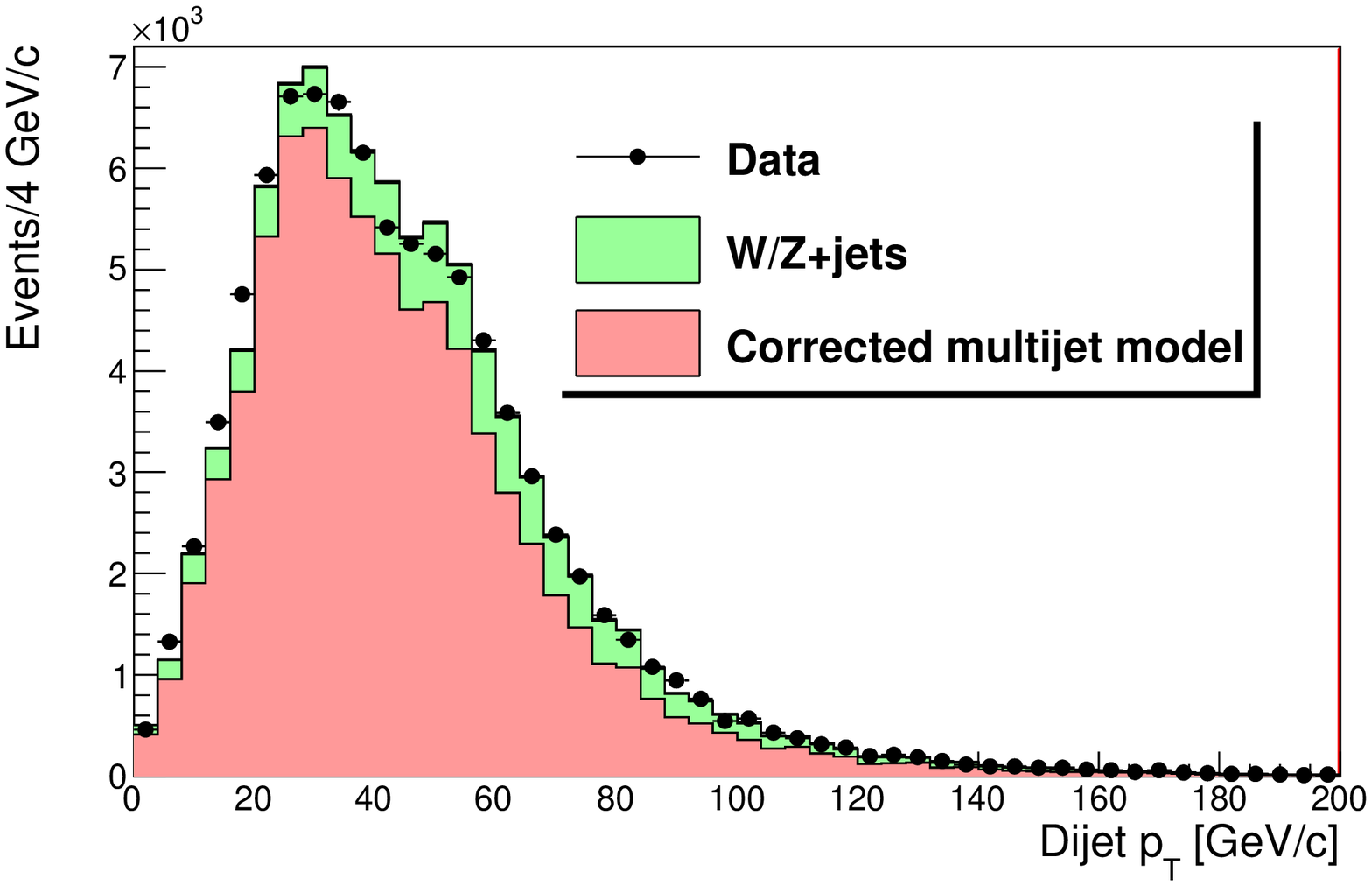}}
\captionsetup{justification=raggedright}
\captionsetup{labelsep=period}
\caption{Transverse-momentum distribution of the two-jet system in the multijet-enriched control sample as observed in the data (circles) and as predicted by the $W/Z+$jets simulation (light shaded histogram) and the nonelectron-based model (dark shaded histogram) before (a) and after (b) application of tunings to the nonelectron-based multijet model. The magnitude of $W/Z$+jets contributions is normalized to the NLO calculations \cite{VjetsNorm}, while the magnitude of the multijet model is obtained from the data.}
\label{fig:dijetptsideband}
\end{figure*}

In order to test the tunings, we use the multijet-enriched control region. An important kinematic distribution related to the dijet-invariant mass is the $p_T$ of the two-jet system. Figure\ \ref{fig:dijetptsideband} shows the improvement in the modeling of this variable after all tunings are applied and is indicative of the improvement seen in other relevant kinematic variables.

\begin{figure*}[htbp]
 \centering 
\captionsetup{justification=centering}
\subfloat[] {\includegraphics[width=0.49\textwidth]{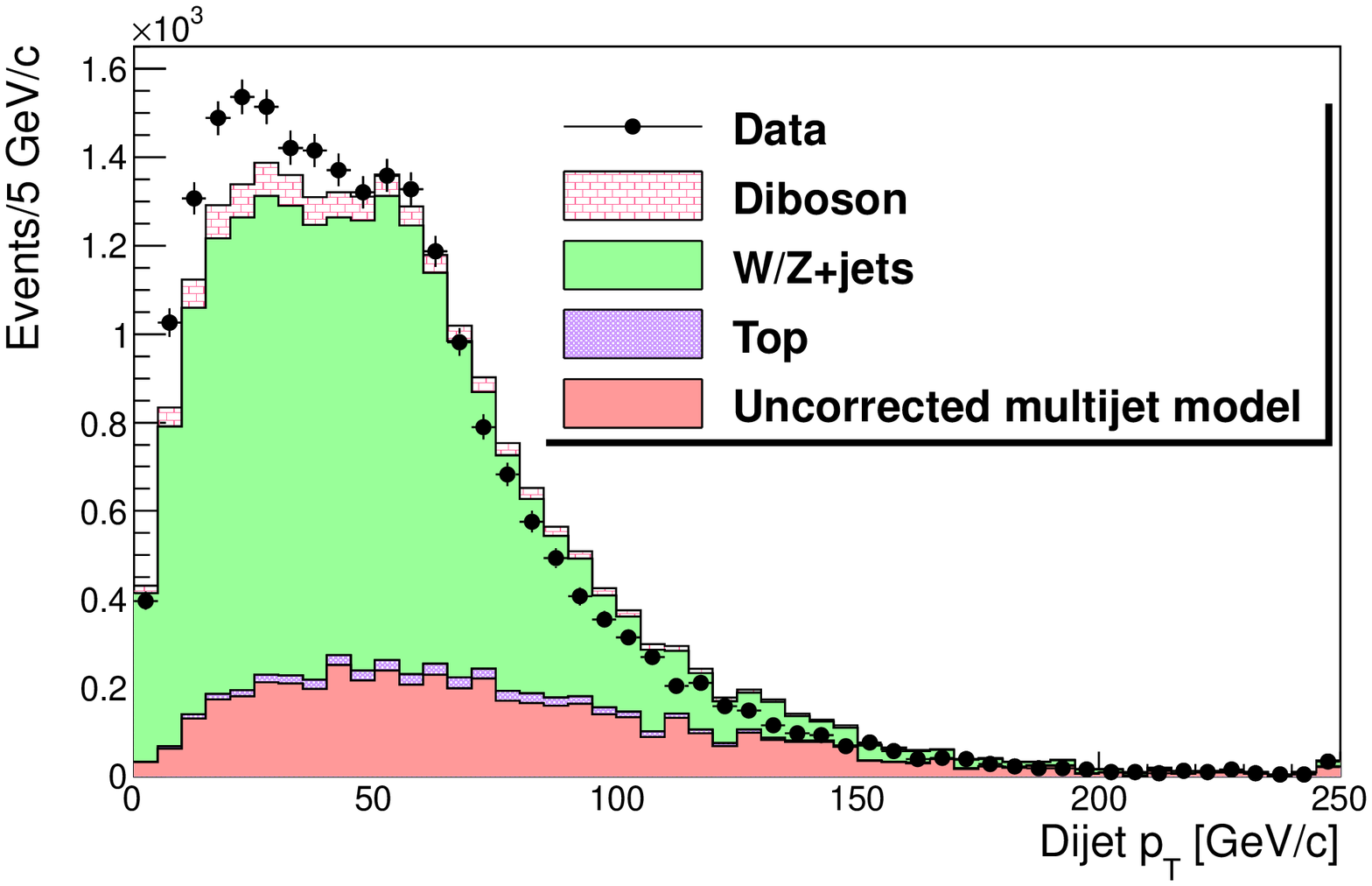}}
\subfloat[] {\includegraphics[width=0.49\textwidth]{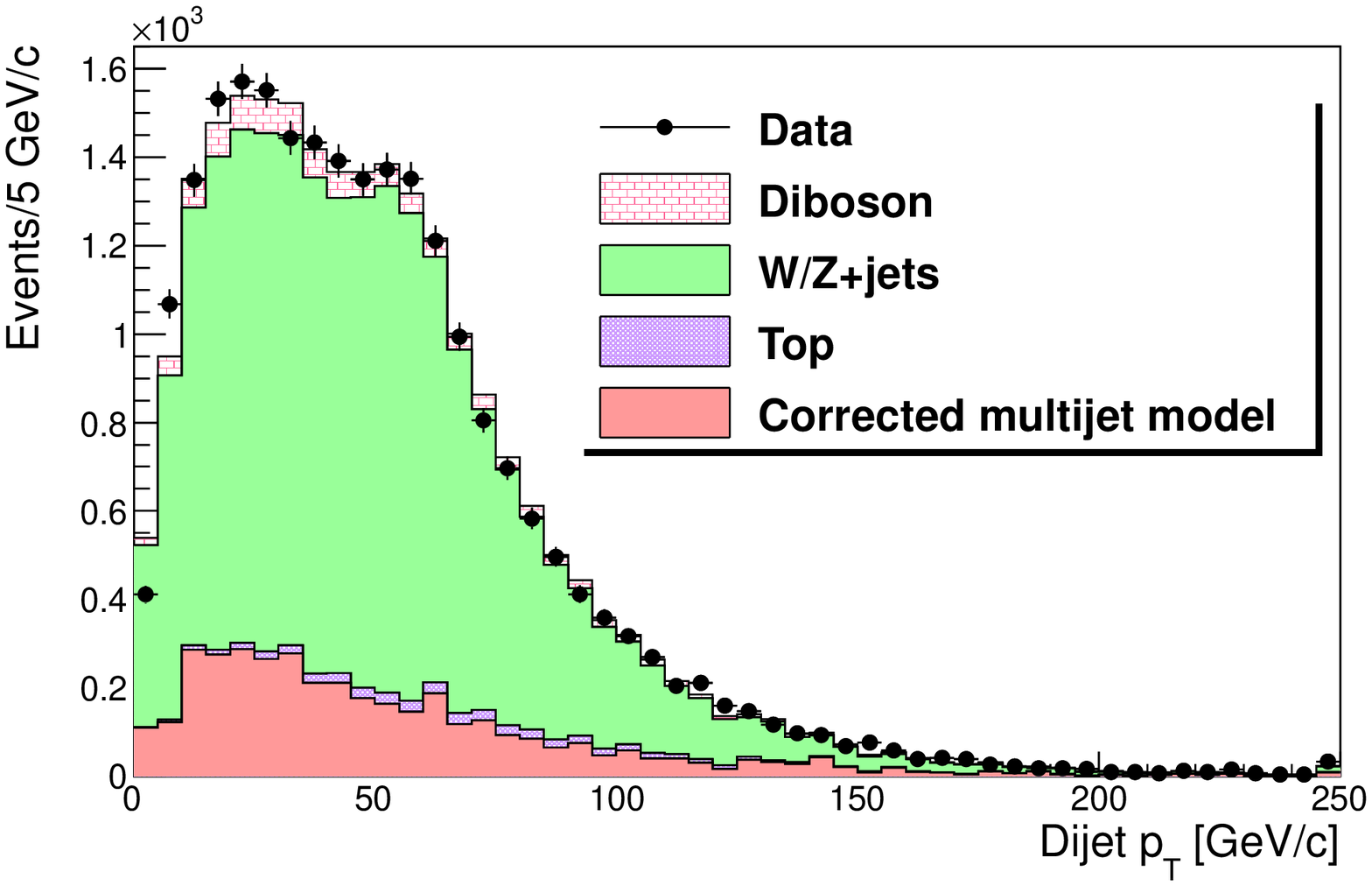}}
\captionsetup{justification=raggedright}
\captionsetup{labelsep=period}
 \caption{Transverse-momentum distribution of the two-jet system in the selected event sample with looser selection criteria as observed in the data and as predicted by the models before (a) and after (b) application of tunings to the nonelectron-based multijet model. }
 \label{fig:ptdibosons}
 \end{figure*}

We also investigate the impact of the tunings applied to the nonelectron-based multijet model on the signal sample, defined in Sec.~\ref{sec:selection}.  To increase the statistical accuracy of the sample, we loosen the selection by removing the two-jet system $p_T$ requirement and lowering the $E_T$ requirements to 25\,GeV. The resulting improvement in the modeling of the two-jet system $p_T$ distribution in this sample is shown in Fig.\ \ref{fig:ptdibosons}.

\begin{figure*}[htbp]
\centering 
\captionsetup{justification=centering}
\subfloat[] {\includegraphics[width=0.49\textwidth]{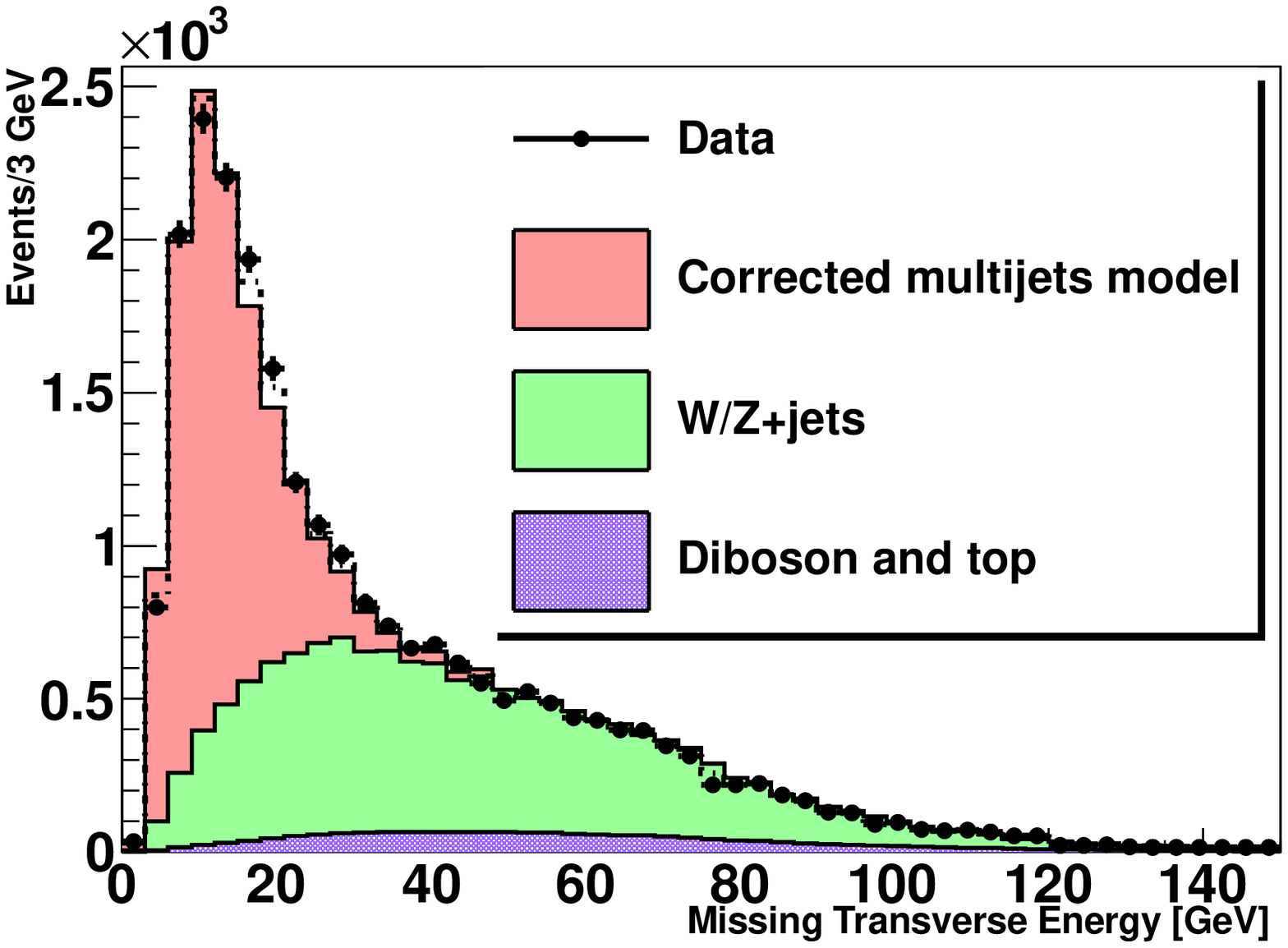}}
\subfloat[] {\includegraphics[width=0.49\textwidth]{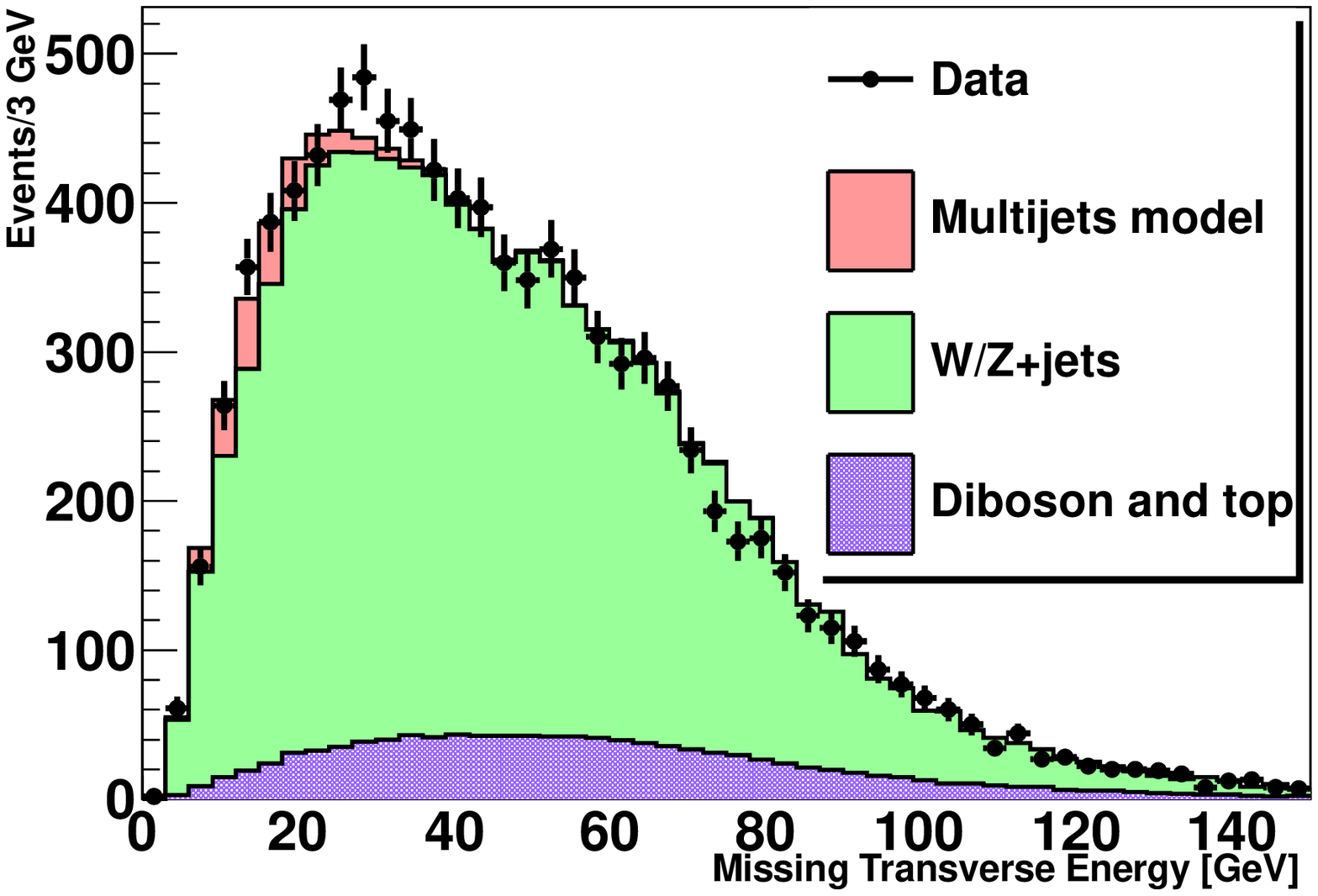}}
\captionsetup{justification=raggedright}
\captionsetup{labelsep=period}
\caption{Missing transverse-energy distribution for events containing electrons (a) and muons (b) from the selected sample. The distributions of observed data are shown with fit background overlaid.}
\label{fig:qcdmetfit}
\end{figure*}

The contribution of the multijet background to the selected sample is determined using a three-component fit to the \met distribution in the data. The three components are the multijet background, the $W/Z$+jets production, and the other electroweak processes (top-quark and diboson production). The last component is constrained to theoretical predictions, whereas the magnitudes of the $W/Z$+jets and the multijet contributions are allowed to float in the fit. The results are shown in Fig.\ \ref{fig:qcdmetfit}. We estimate the amount of multijet background in the electron and muon sample to be $(7.8 \pm 0.2)\%$ and $(0.27 \pm 0.01)\%$ respectively, where the uncertainties are statistical only. We consider several systematic uncertainties: jet-energy-scale modeling ($0.9\%$), choice of the fit variable ($13.1\%$), disagreement between the observed and predicted multijet \met distribution ($4.4 \%$), and theoretical uncertainties on the cross sections ($0.9\%$). The total systematic uncertainty on the multijet background estimate is $14.0\%$.

%%% Local Variables: 
%%% mode: latex
%%% TeX-master: "DijetMassSpectra.tex"
%%% End: 

\section{Fit and Results}
\label{sec:FitResults}
We first describe the procedure used to fit the
observed dijet-mass distribution in data, including contributions from
background and an hypothetical signal. We then present two sets of results. For
the first set, we do not incorporate the specific jet-energy-scale
corrections for quark and gluon jets nor the tuning of the
multijet-background model, essentially performing the analysis of Ref.\
\cite{bumpprl} on the full CDF Run II data set. The final
results are then given, which include those
obtained when the improvements are incorporated.

\subsection{Fit technique}
\label{sec:Fit_Techniques}
Uncertainties on the predictions are parametrized
with nuisance parameters, and the data are used to constrain both the
signal size and the values of these parameters.

We use the following approach to set an upper limit on the production
rate of a hypothetical new particle.  We maximize a binned likelihood function 
$L({\rm{data}}|{\vec{\theta}},{\vec{\nu}})\pi({\vec{\nu}})$, which expresses
the probability of observing the data given the model parameters ${\vec{\theta}}$ and the nuisance parameters 
${\vec{\nu}}$.  The likelihood is a product of Poisson probabilities
for the observed data in each bin. The
function $\pi({\vec{\nu}})$ is a product of Gaussian constraints, one
for each systematic uncertainty (treated as nuisance parameters in the fit), which incorporates
external information about the parameter, as measured in control
samples or obtained from other sources.
The nuisance parameters describe three classes of systematic
uncertainties: bin-by-bin uncertainties, which are considered
uncorrelated between individual bins of each predicted distribution;
shape uncertainties, which correspond to coherent
distortions across the bins of a distribution, parametrized by a single nuisance parameter; and rate
uncertainties, which coherently affect the normalization of all bins
within one distribution.  Rate and shape uncertainties
may be correlated. For example, modifications of the jet-energy-scale
shift the mass of a resonance to higher or lower values (Fig.\ \ref{fig:jeshiVsjesloVsnominalWjetsBumpCuts}); in
addition, they affect the magnitude of the predicted contribution of the
process due to the selection criterion that jets pass a
minimum $E_{\rm T}$ threshold.  These correlations are taken into account
by allowing each source of systematic uncertainty to affect both rates and shapes of multiple distributions.  A detailed description of the likelihood function
is given in Ref.~\cite{Aaltonen:2010jr}.  Restrictions are placed on the allowed ranges of the nuisance parameters
to ensure that all event-yield predictions are non-negative.
%The program {\sc minuit}~\cite{James:1975dr} is used to perform the
%maximization of the likelihood function. The associated uncertainties
%are computed using the {\sc minos} routine of {\sc minuit}.  

\subsection{Results}
\label{sec:Results}
\begin{figure*}[htbp]
\centering
{\includegraphics[width=0.79\textwidth]{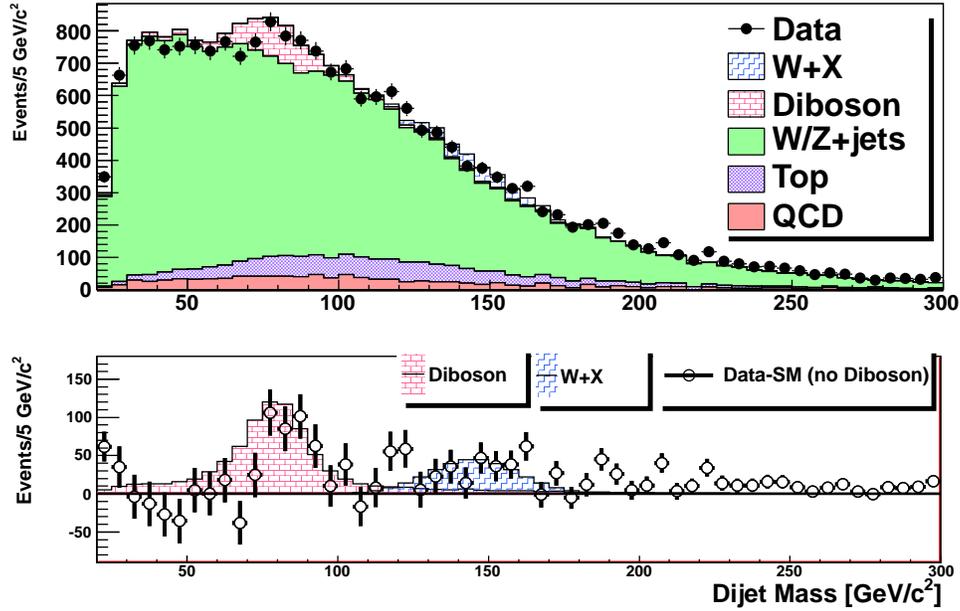}}
\captionsetup{labelsep=period}
\caption{Dijet mass distribution with fit results overlaid in the combined electron and muon data sets prior to incorporating the improvements
  discussed in the text, equivalent to updating the analysis described
  in Ref.\ \cite{bumpprl} on the full CDF data set. The bottom panel shows
  data with all fit background contributions subtracted except those from diboson production.}
\label{fig:noimprovements}
\end{figure*}

\begin{figure*}[htbp]
\centering 
\captionsetup{justification=centering}
\subfloat[] {\includegraphics[width=0.49\textwidth]{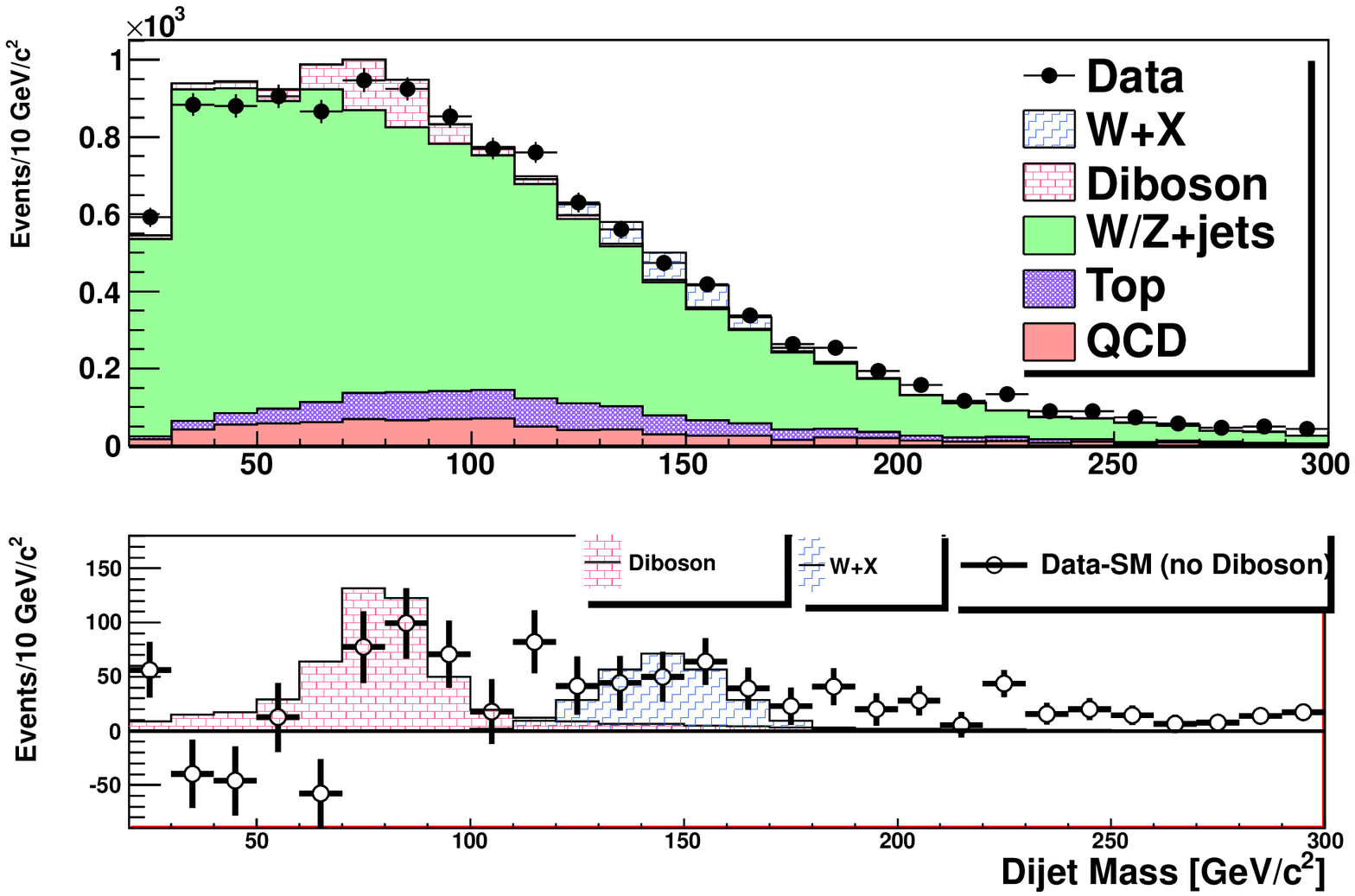}}
\subfloat[] {\includegraphics[width=0.49\textwidth]{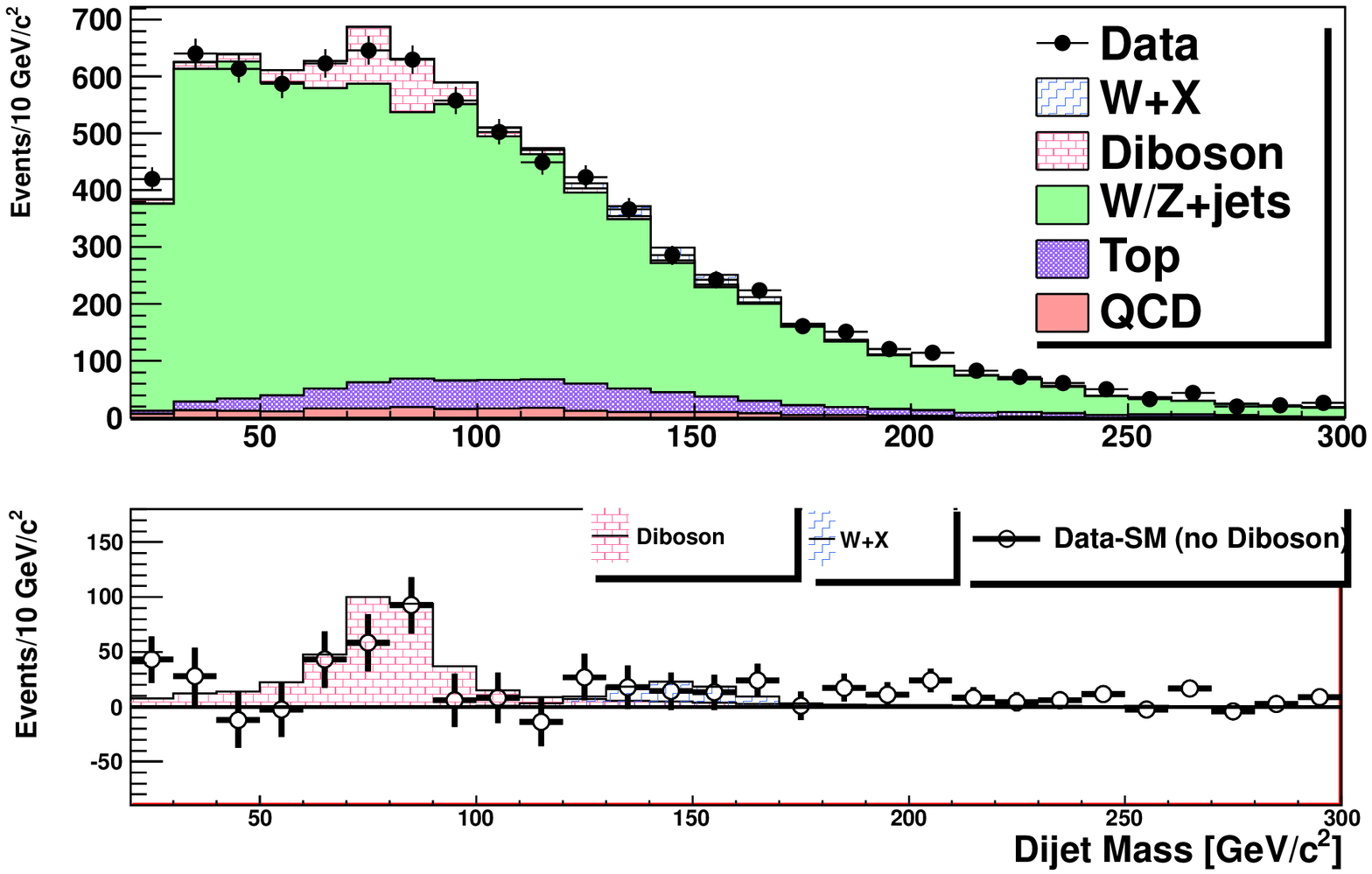}}
\captionsetup{justification=raggedright}
\captionsetup{labelsep=period}
\caption{Same distribution as in Fig. \ref{fig:noimprovements}, shown separately for the electron (a) and muon (b) samples.}
\label{fig:noimpelectronmuons}
\end{figure*}

To reproduce the previous analysis\ \cite{bumpprl}, a first
fit to the dijet invariant-mass spectrum is performed without
incorporating the improvements described in the previous sections. In addition
to the SM contributions, an additional Gaussian component centered at $144~{\rm GeV}/c^2$ with a
width of $14.3~{\rm GeV}/c^2$ is incorporated in the fit to model a
potential non-SM contribution. The result of the fit in the full
electron and muon data sample is shown in Fig.\
\ref{fig:noimprovements}: an excess of events over the background
prediction is observed in the signal region, similar to what observed
in Ref.\ \cite{bumpprl}. Assuming that this new contribution has
the same acceptance as that for a 140 GeV/c$^2$ Higgs boson produced in association with a $W$ boson,
the extracted cross section is 2.4 $\pm$ 0.6 pb. Assuming only
SM processes, the probability to measure a value as large or larger than the observed cross section is $2.6\times10^{-5}$, which corresponds to $4.2\sigma$ in terms of standard
deviations. The excess is similar in the electron and muon channels,
as shown in Fig.\ \ref{fig:noimpelectronmuons}.

\begin{figure*}[htbp]
\centering
{\includegraphics[width=.79\textwidth]{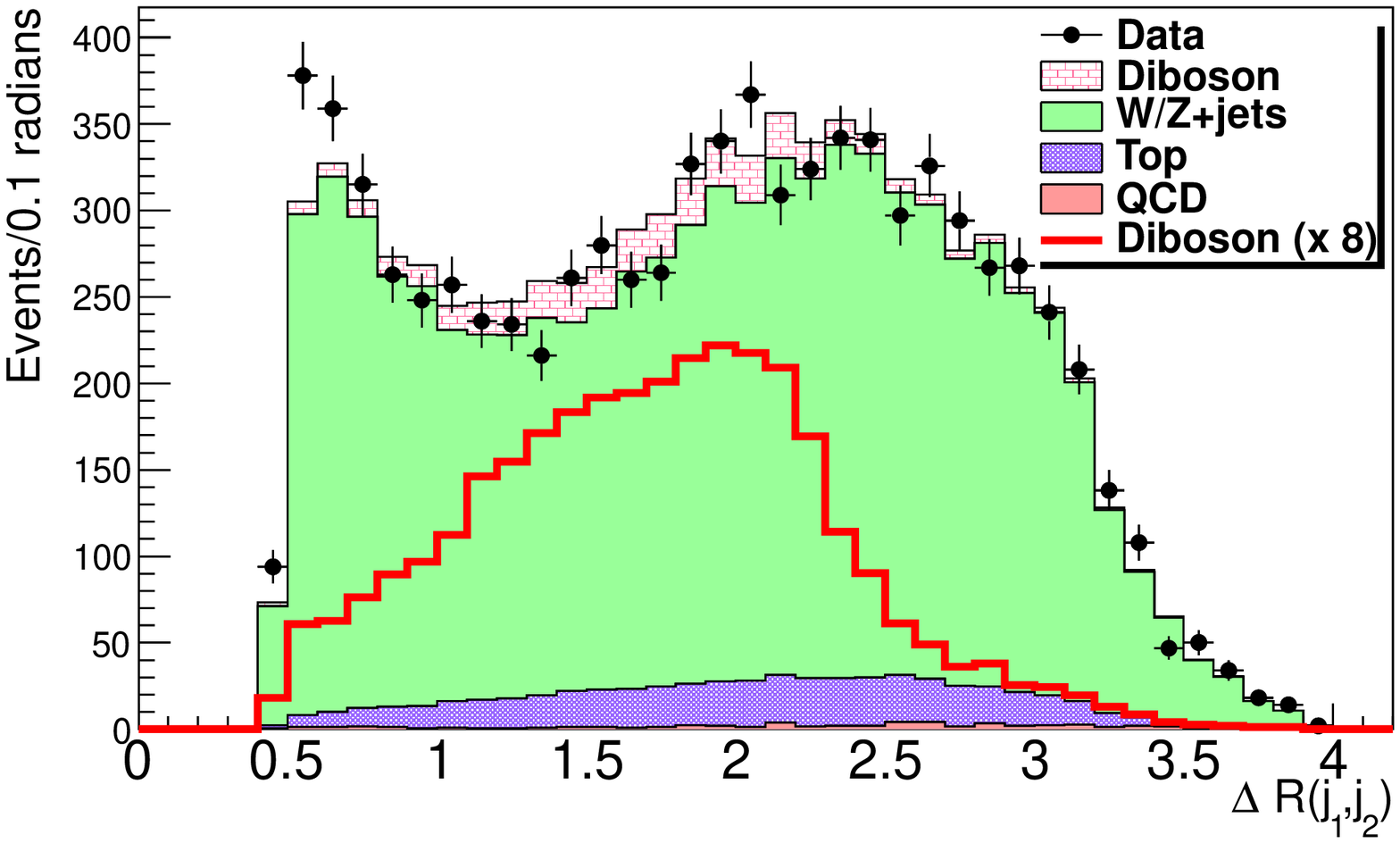}}
\captionsetup{labelsep=period}
\caption{Distribution $\Delta R(j_1,j_2)$ distributions in the muon sample as
  observed in the data and as predicted by the models incorporating
  improved jet-energy-scale corrections for simulated quark and gluon
  jets. The diboson distribution (red line) magnified by a factor of 8
  is also shown as a example of the $\Delta R(j_1,j_2)$ distribution for a heavy-particle decay.}
\label{fig:dRbumpcutsMuosample}
\end{figure*}

\begin{figure*}[htbp]
{\includegraphics[width=.79\textwidth]{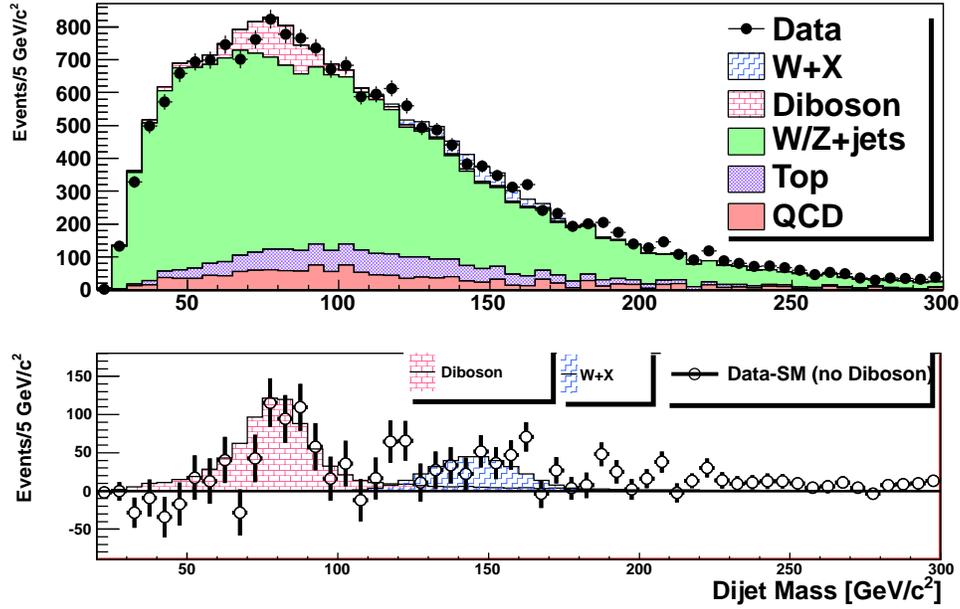}}
\captionsetup{labelsep=period}
\caption{Dijet mass distribution with fit results overlaid in the
  combined electron and muon data sets selected by applying an
  additional $\Delta R(j_1,j_2)>0.7$ requirement and prior to incorporating the improvements
  discussed in the text. The bottom panel shows
  data with all fit background contributions subtracted except
  those from diboson production.}
\label{fig:noimprovements_otherthandR07cut}
\end{figure*}

\begin{figure*}[htbp]
\captionsetup{justification=centering}
\subfloat[] {\includegraphics[width=.49\textwidth]{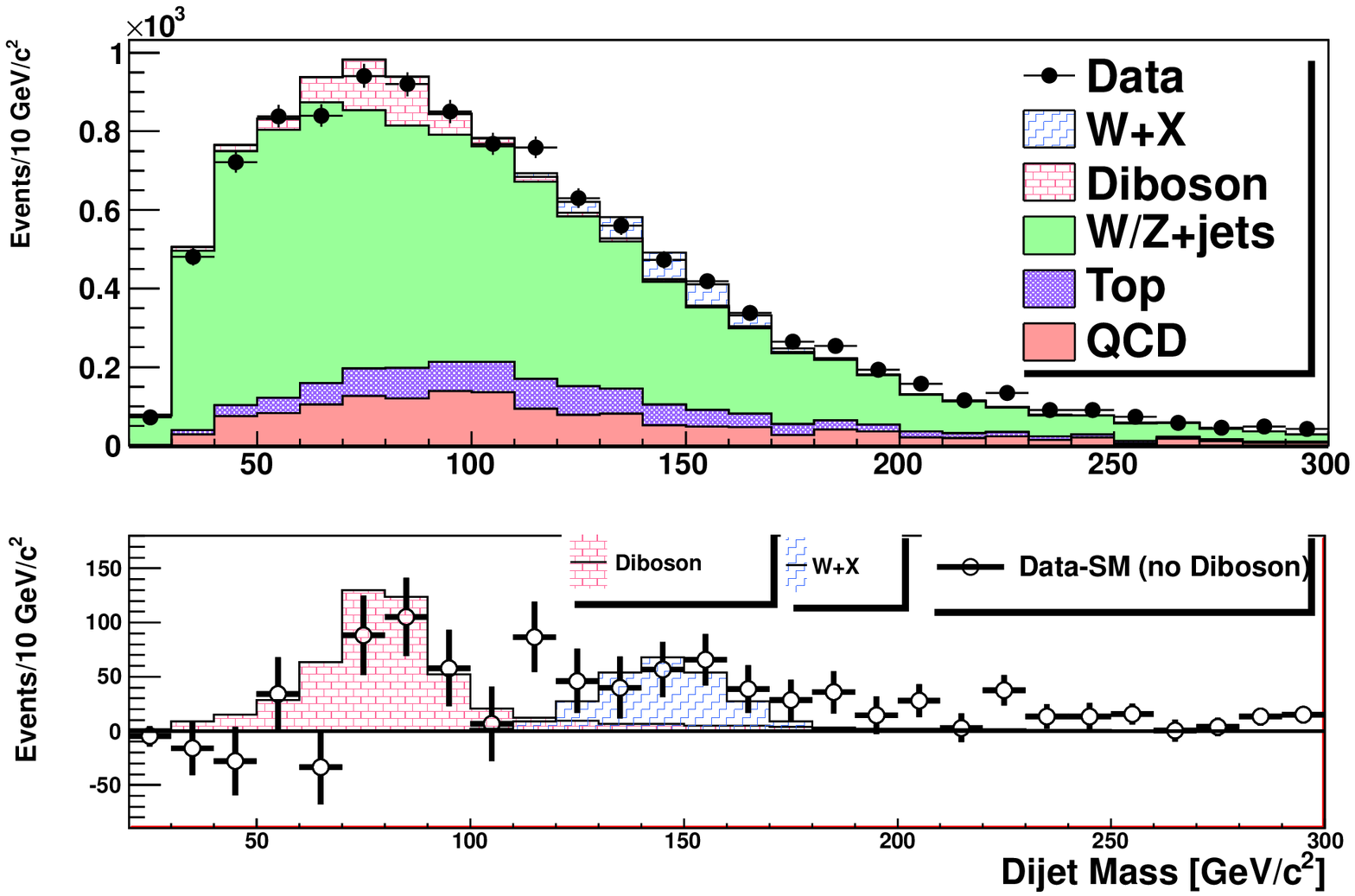}}
\subfloat[] {\includegraphics[width=.49\textwidth]{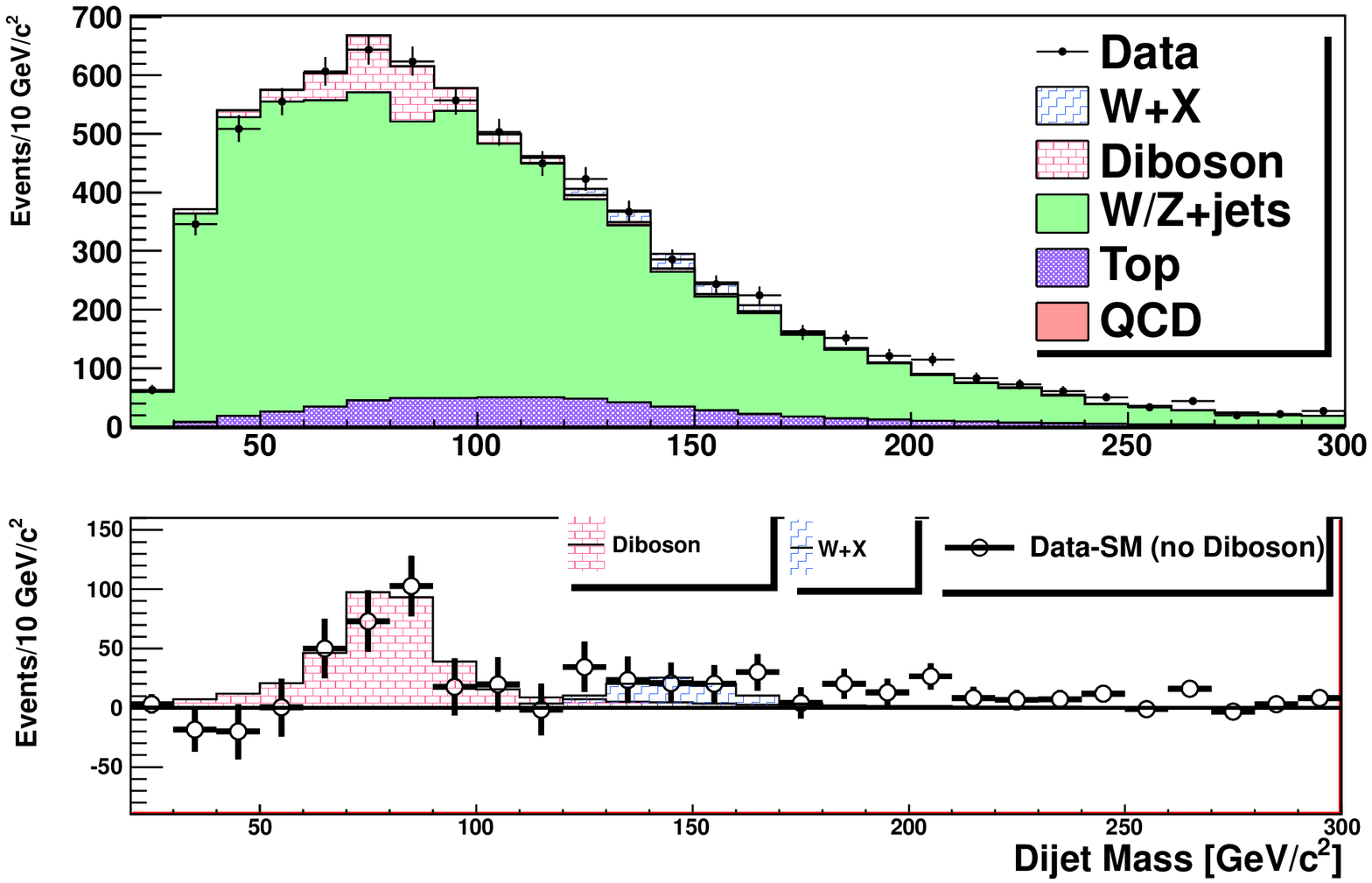}}
\captionsetup{justification=raggedright}
\captionsetup{labelsep=period}
\caption{Same distribution as in Fig.\ \ref{fig:noimprovements_otherthandR07cut}, shown separately for the electron (a) and muon (b) samples.}
\label{fig:noimprovements_otherthandR07cutelemuo}
\end{figure*}

Figure~\ref{fig:dRbumpcutsMuosample} shows that the SM predictions do
not model properly the region at low $\Delta R$ between the two jets
($\Delta R(j_1,j_2)$) in the muon sample. A similar discrepancy is
observed in the electron sample. However, jet
pairs from heavy particles are expected to be
produced more often at large $\Delta R(j_1,j_2)$. Therefore, applying a $\Delta
R(j_1,j_2)>0.7$ requirement is not expected to bias heavy-particle searches. Nonetheless, we investigate the effect of this requirement on the
final
result. Figures \ref{fig:noimprovements_otherthandR07cut}-\ref{fig:noimprovements_otherthandR07cutelemuo}
show that, although the agreement between data and SM expectations in
the region at low masses is improved, similar discrepancies as in
Figs. \ref{fig:noimprovements}-\ref{fig:noimpelectronmuons} are
present for dijet-invariant masses larger than 50 GeV/$c^2$. We
extract a cross section $\sigma_{\it WX} = (2.3 \pm 0.5)$ pb, which is
compatible with the one extracted with no $\Delta R(j_1,j_2)$ restriction. 

\begin{figure*}[htbp]
\centering 
\captionsetup{justification=centering}
\subfloat[] {\includegraphics[width=0.49\textwidth]{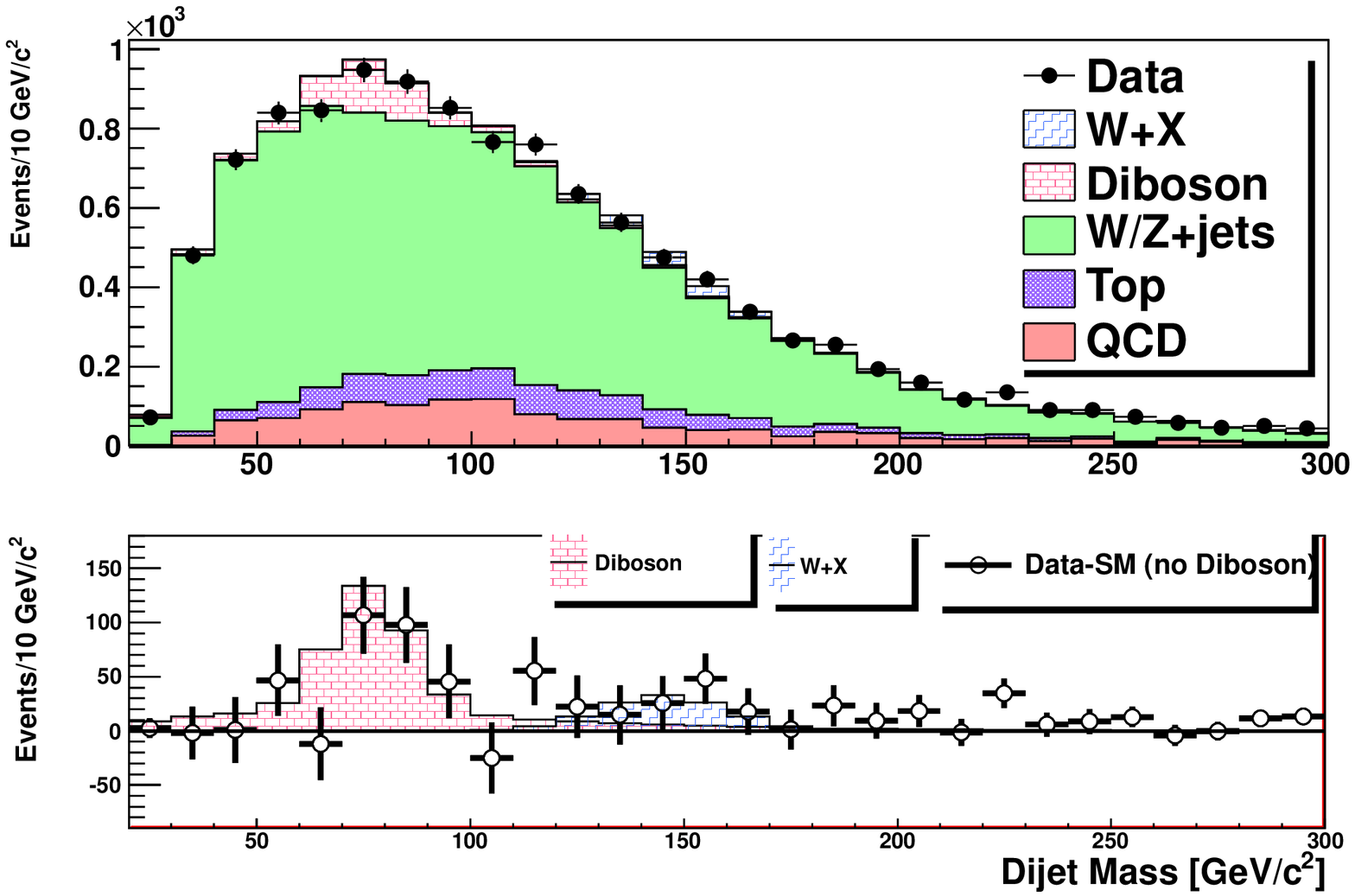}}
\subfloat[] {\includegraphics[width=0.49\textwidth]{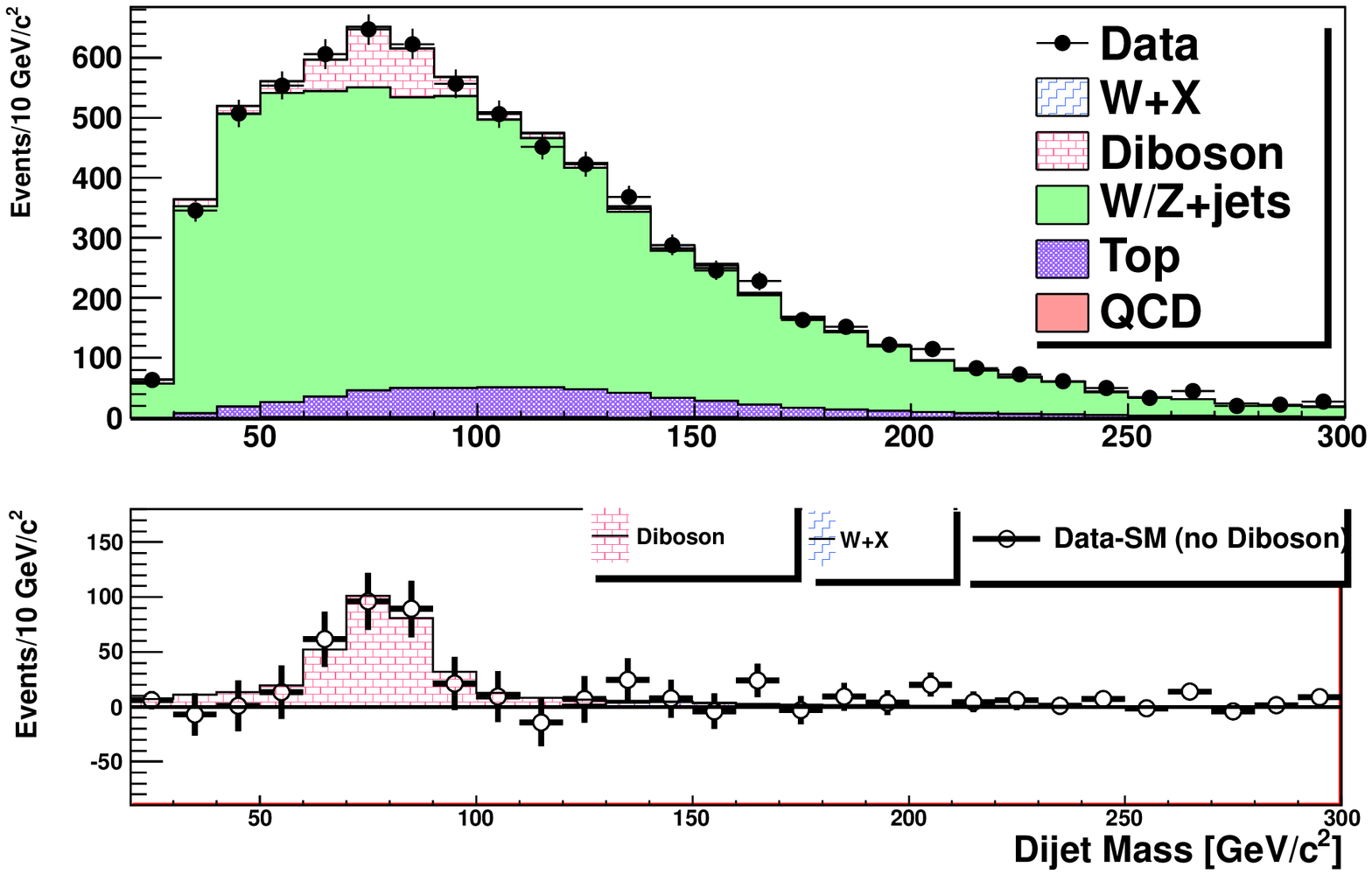}}
\captionsetup{justification=raggedright}
\captionsetup{labelsep=period}
\caption{Dijet mass distribution with fit results overlaid in the electron (a) and muon data sets (b) incorporating improved jet-energy-scale corrections for
  simulated quark and gluon jets but no tuning on the multijet-background modeling. The bottom panel shows
  data with all fit background contributions subtracted except those from diboson production.}

\label{fig:qgafter}
\end{figure*}

\begin{figure*}[htbp]
\centering
{\includegraphics[width=0.49\textwidth]{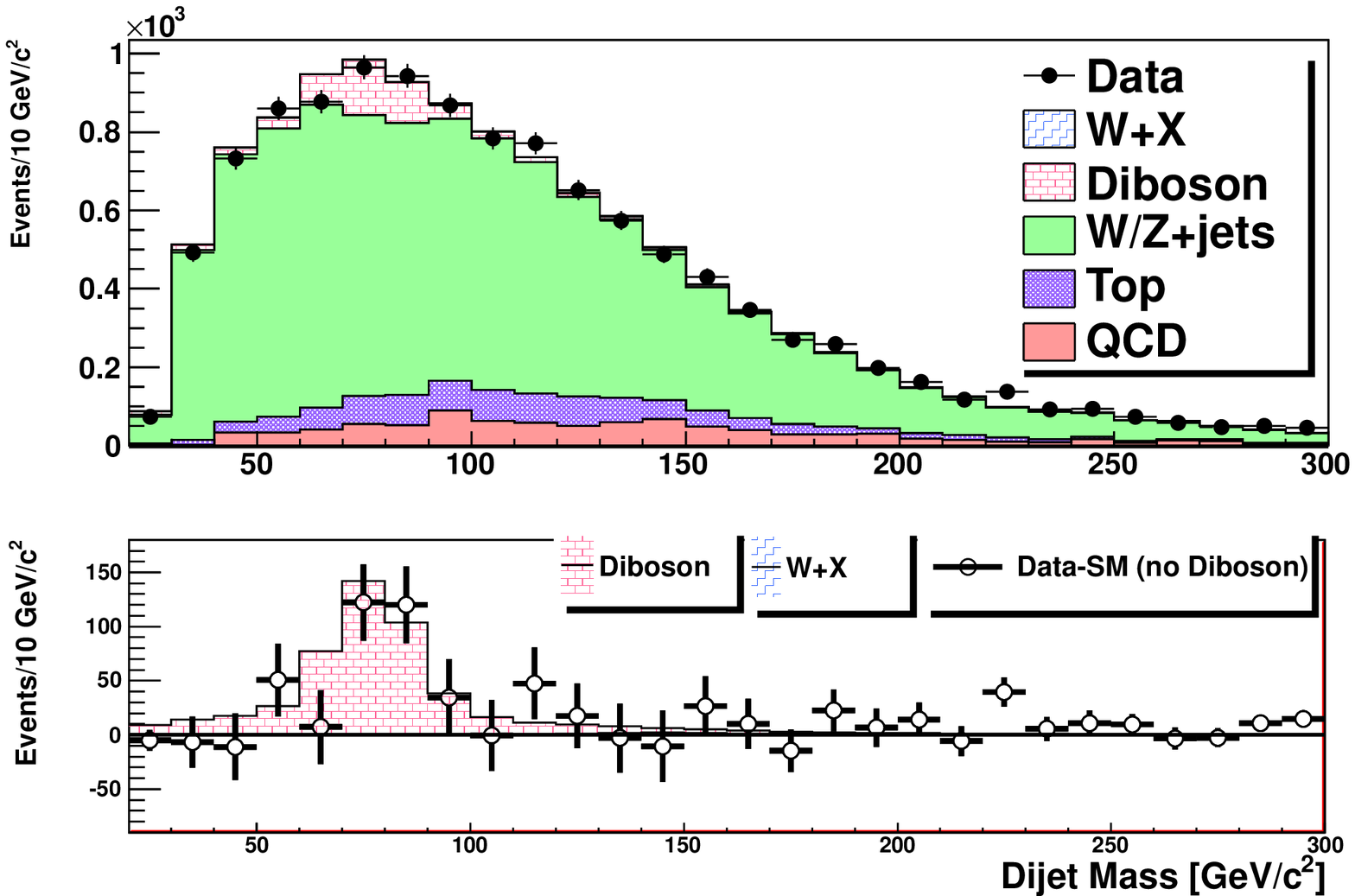}}
\captionsetup{labelsep=period}
\caption{Dijet mass distribution with fit results overlaid in the electron data set incorporating improved jet-energy-scale corrections for
  simulated quark and gluon jets and tunings on the multijet-background modeling. The bottom panel shows
  data with all fit background contributions subtracted except those from diboson production.}
\label{fig:elewithimprovements}
\end{figure*}

Additional fits incorporate the corrections described in 
Secs.~\ref{sec:JESModeling} and~\ref{sec:mult-backgr-model}. First, jet-energy-scale corrections for simulated quark and gluon
jets described in Sec.~\ref{sec:JESModeling} are incorporated. The
resulting fits to the selected-event distributions with electrons and muons are
shown separately in Fig.\ \ref{fig:qgafter}. Good agreement between the observed
data and the fit contributions is seen for events with muons, while the agreement is
still rather poor for events with electrons. Final fits performed
after incorporating also tunings to the multijet-background model lead
to excellent agreement between the observed electron data and the
fit-SM-process contributions, as shown in
Fig.\ \ref{fig:elewithimprovements}. The fit to the muon data, where the multijet background is very small, is unchanged.

\begin{figure*}[htbp]
\centering
{\includegraphics[width=0.79\textwidth]{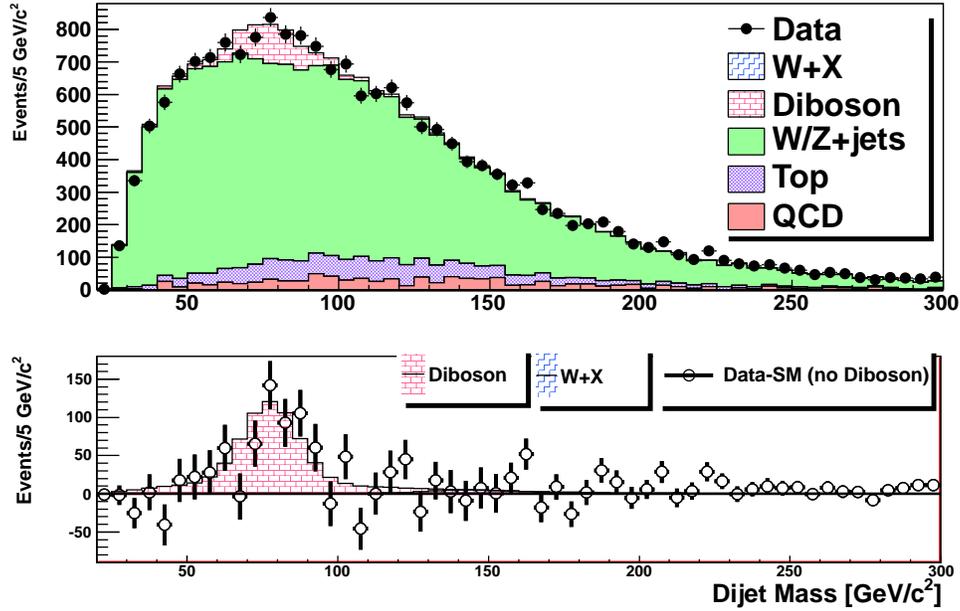}}
\captionsetup{labelsep=period}
\caption{Dijet mass distribution with fit results overlaid in the combined electron and muon data sets incorporating improved jet-energy-scale corrections for
  simulated quark and gluon jets and tunings on the multijet-background modeling. The bottom panel shows
  data with all fit background contributions subtracted except those from diboson production.}
\label{fig:withimprovements}
\end{figure*}

The final fit result for the combined electron and muon data is shown
in Fig.\ \ref{fig:withimprovements}. The magnitude of SM contributions
is normalized to the expected rates given in Table \ref{tab:RatesBumpCuts}. Since the data
are consistent with the SM predictions and no significant excess is
observed, we set an upper limit
of $0.9$ pb at the 95$\%$ C.L. on the cross section of a new particle
with a mass of 144 GeV/$c^2$ produced in association with a $W$
boson. The limit assumes that the new resonance has an acceptance equal to
that of a Higgs boson produced in association with a $W$ boson, and
the limit is set using likelihood-ratio ordering \cite{FCbands}. When
generating pseudoexperiments we start from the rates in Table
\ref{tab:RatesBumpCuts} and we allow for variations within systematic
uncertainties mentioned in Sec.\
\ref{sec:SignalBackgroundModeling}. Shape variations due to the
jet-energy-scale, factorization and normalization scale
uncertainties are also considered. %where we analyze the distribution of measured cross
%sections in pseudoexperiments generated with a variety of scale
%factors (0.1 to 1.2 at a step of 0.1) on the input cross
%section. Fig. \ref{fig:FCBands} shows the results of the
%Feldman-Cousins analysis. The esablished upper limit is
%$0.9$ pb at the 95$\%$ C.L. 

\begin{comment}
\begin{figure}[htbp]
\centering
{\includegraphics[width=.5\textwidth]{FCBands.eps}}
\caption{Feldman-Cousins bands showing the expected range of measured cross sections as
a function of the true cross section, with 68$\%$ CL (light blue region) and 95$\%$ CL (dark blue region). The measured cross-section is (0.0 $\pm$ 0.7) pb. %http://www-cdf.fnal.gov/internal/WebTalks/Archive/1212/121221_higgs/02_121221_higgs_Marco_Trovato_1_mtrovato_Dec21HiggsMeeting.pdf
The intersection of the measured cross-section and the 95$\%$ CL bands corresponds to the upper limit $\sigma_{WX} <$ 0.9 pb at the 95$\%$ C.L.}
\label{fig:FCBands}
\end{figure}
\end{comment}

%%% Local Variables: 
%%% mode: latex
%%% TeX-master: "DijetMassSpectra.tex"
%%% End: 

%\section{Systematic Uncertainties} %Pending... \include{Systematic_Uncertainties}

%\section{$W + jj$~Shape} %Pending... \include{Wjj_Shape}

%\includeorinput{invariantmassspectrum}

%\includeorinput{othersamples}

\section{Conclusion}
\label{sec:conclusion}
We present a study of the dijet invariant-mass spectrum in events containing a single lepton, large missing transverse energy, and exactly two jets. Since the previous publication \cite{bumpprl}, additional studies of potential systematic effects have led to the incorporation of specific jet-energy-scale corrections for simulated quark and gluon jets and tunings of the data-driven modeling for the multijet-background contributions. The distribution observed in the full CDF Run II data set is in good agreement with the SM expectations, whose dominant contributing process is $W+$jets, which is modeled using {\sc alpgen} event generator combined with {\sc pythia} simulation of parton showering and hadronization. A 95$\%$ C.L. upper limit of 0.9 pb is set on the cross section times branching ratio for production and decay into dijets of a new particle with mass of 144 GeV/$c^2$ in association with a $W$ boson.

\newpage
\begin{acknowledgments}
We thank the Fermilab staff and the technical staffs of the
participating institutions for their vital contributions. This work
was supported by the U.S. Department of Energy and National Science
Foundation; the Italian Istituto Nazionale di Fisica Nucleare; the 
Ministry of Education, Culture, Sports, Science and Technology of
Japan; the Natural Sciences and Engineering Research Council of
Canada; the National Science Council of the Republic of China; the
Swiss National Science Foundation; the A.P. Sloan Foundation; the
Bundesministerium f\"ur Bildung und Forschung, Germany; the Korean
World Class University Program, the National Research Foundation of
Korea; the Science and Technology Facilities Council and the Royal
Society, UK; the Russian Foundation for Basic Research; the
Ministerio de Ciencia e Innovaci\'{o}n, and Programa
Consolider-Ingenio 2010, Spain; the Slovak R\&D Agency; the Academy
of Finland; the Australian Research Council (ARC); and the EU
community Marie Curie Fellowship contract 302103.

\end{acknowledgments}

\newpage

%%% Local Variables: 
%%% mode: latex
%%% TeX-master: "DijetMassSpectra.tex"
%%% End: 

%\bibliography{DijetMassSpectra}

\end{document}